\RequirePackage{ifpdf}

\documentclass[a4paper,11pt]{article}
\pdfoutput=1 

\usepackage{jheppub} 

\usepackage[T1]{fontenc} 

\usepackage{ifpdf} 

\usepackage{flexisym}
\usepackage{breqn}
\usepackage{comment}

\usepackage{epsf,epsfig,epstopdf}
\usepackage{graphics}
\usepackage{graphicx}
\usepackage[usenames,dvipsnames]{xcolor}
\definecolor{darkgreen}{RGB}{34,139,34}

\usepackage{enumerate}

\usepackage{amsmath,amsfonts,amssymb}

\newcommand{\id}{\mathrm{d}}
\newcommand{\nn}{\nonumber}
\newcommand{\eps}{\epsilon}
\newcommand{\define}{\equiv}
\newcommand{\tekst}{\textrm}
\newcommand{\kur}[1]{\mathcal{#1}}
\newcommand{\Li}{\tekst{Li}}
\newcommand{\sgn}{\tekst{sgn}}
\newcommand{\Litt}{\Li_{2,2}}
\newcommand{\floor}[1]{\left \lfloor #1 \right \rfloor}
\newcommand{\bin}[2]{\left( \!\! \begin{array}{c} #1 \\ #2 \end{array} \!\! \right)}
\newcommand{\Max}{\tekst{Max}}
\newcommand{\scl}[1]{\scalebox{0.6}{\(#1\)}}

\DeclareFontFamily{U}{wncy}{}
\DeclareFontShape{U}{wncy}{m}{n}{<->wncyr10}{}
\DeclareSymbolFont{mcy}{U}{wncy}{m}{n}
\DeclareMathSymbol{\sha}{\mathord}{mcy}{"58}

\title{On the reduction of generalized polylogarithms to $\Li_n$ and $\Litt$ and on the evaluation thereof}

\author[a]{Hjalte Frellesvig}
\author[a]{Damiano Tommasini}
\author[a,b,c]{Christopher Wever}
\affiliation[a]{
Institute of Nuclear and Particle Physics, NCSR ``Demokritos'',\\
Patriarchou Gregoriou E. \& Neapoleos 27, Agia Paraskevi, 15310, Greece
}
\affiliation[b]{
Institute for Theoretical Particle Physics (TTP), Karlsruhe Institute of Technology,\\
Engesserstra{\ss}e 7, D-76128 Karlsruhe, Germany
}
\affiliation[c]{
Institute for Nuclear Physics (IKP), Karlsruhe Institute of Technology,\\
Hermann-von-Helmholtz-Platz 1, D-76344 Eggenstein-Leopoldshafen, Germany
}

\emailAdd{frellesvig@inp.demokritos.gr, tommasini@inp.demokritos.gr, christopher.wever@kit.edu}

\preprint{TTP16-002}

\abstract{%
We give expressions for all generalized polylogarithms up to weight four in terms of the functions $\log$, $\Li_n$, and $\Litt$, valid for arbitrary complex variables. Furthermore we provide algorithms for manipulation and numerical evaluation of $\Li_n$ and $\Litt$, and add codes in Mathematica and C++ implementing the results. With these results we calculate a number of previously unknown integrals, which we add in App. \ref{app:integrals}.
}
\notoc

\begin{document}
\maketitle
\flushbottom


\section{Introduction}
\label{sec:intro}

In the recent decades, the Standard Model of particle physics has been established as extremely successful, due to its high level of agreement with the data provided by numerous experiments.
In order to further test the Standard Model, or possibly discover new physics, we need to both predict and measure the values of many observables with even higher accuracy.

From the theoretical side, which is the focus of this paper, this necessitates the computation of higher order corrections in perturbation theory, and for that purpose new mathematical techniques to compute loop integrals become desirable.
A large class of dimensionally regularized Feynman loop-integrals can be expressed in terms of generalized (or Goncharov) polylogarithms (GPLs) (in the mathematical community these are also known as hyperlogarithms).
Those special functions were originally introduced by Kummer \cite{Kummer:1840} and Poincar\'e \cite{Poincare:1883}, then further studied by Chen \cite{chen1977}, and brought to the attention of the physics community by Goncharov \cite{Goncharov:2011ar, Goncharov:2010jf}.

At one-loop all Feynman integrals (in $4-2 \eps$ dimensions) are expressible in terms of the logarithm $\log(x)$ and the dilogarithm $\Li_2(x)$, up to the zeroth order in the $\epsilon$ expansion \cite{Ellis:2007qk}, and these functions are special cases of GPLs.
At two or more loops many Feynman integrals can be likewise expressed in terms of GPLs \cite{Remiddi:1999ew, Gehrmann:2000zt, Aglietti:2004tq,Korner:2005qz, Bonciani:2010ms,Ablinger:2011te,Henn:2014lfa,Caola:2014lpa,Papadopoulos:2014hla,Ablinger:2014nga,Gehrmann:2015bfy,Papadopoulos:2015jft,Henn:2013pwa} (for further references, see \cite{Duhr:2014woa,Henn:2014qga} and the references therein), but there are also integrals which are counter examples, such as notably that of the fully massive sunset graph \cite{Laporta:2004rb,Adams:2013kgc,Bloch:2013tra,Adams:2014vja,Adams:2015ydq}. Certain graphs without massive propagators are also believed to be counter examples \cite{CaronHuot:2012ab}. In this paper we will restrict the discussion to GPLs.

In \cite{Duhr:2012po} it was conjectured that all GPLs up to weight four, which includes all GPLs needed for two-loop calculations, can be expressed in terms of logarithms, the classical polylogarithms $\Li_n(x)$ ($n\leq4$), and one extra special function denoted by $\Litt(x,y)$. In the same reference it was explicitly shown that the conjecture holds true for a subset of GPLs, denoted harmonic polylogarithms (HPLs), up to weight four. A number of physical calculations of two-loop Feynman integrals with several scales, e.g. \cite{vonManteuffel:2013uoa,Gehrmann:2015ora,Bonciani:2013ywa,Bonciani:2015eua,Gehrmann:2015dua} have hinted at the truth of that conjecture, and in this paper we will show it explicitly. Additionally we provide a library of all the GPLs up to weight four, containing the reduction to the previously named functions that is valid for arbitrary complex variables.

Since the logarithm and classical polylogarithm are well-known functions, efficient algorithms for their numerical evaluation have been widely studied and developed. On the other hand, the {\tt GiNaC} implementation of Vollinga and Weinzierl \cite{Vollinga:2004sn} is the only publicly available program which can efficiently evaluate the special function $\Litt(x,y)$ for any set of complex arguments\footnote{The function $\Litt(x,y)$ are among the ``two-dimensional HPLs'' which are discussed and implemented in ref. \cite{Gehrmann:2001jv} for some real values of the arguments.}. In this paper we provide an independent algorithm and a numerical code for the efficient evaluation of $\Li_n(x)$ and $\Litt(x,y)$.

Thus the tools for the complete reduction and evaluation of any GPL up to weight four are made available with this paper.

The paper is organized as follows: In sect. \ref{sec:definitionGPLs} we define the GPL functions and review some of their main properties.
In sect. \ref{sec:newrelationsGPLs} we provide some relations that are useful for the reduction of GPLs, and in sect. \ref{sec:gpl} we provide some explicit expressions for the reductions.
In sect. \ref{sec:lin} we describe the Crandall \cite{Crandall} algorithm for the evaluation of $\Li_n(x)$, and propose our modified version.
In sect. \ref{sec:li22} we propose an algorithm to evaluate the $\Litt(x,y)$ function, and in sect. \ref{sec:useofthecode} we describe the added code containing our implementation of the algorithms and reduction. 
Finally in sect. \ref{sec:discussion} we discuss our results and propose some directions for further research.

In appendix \ref{app:appA} we provide some relations for $\Litt(x,y)$, and in app. \ref{app:coefficients} we provide details on certain expansions for $\Litt(x,y)$. In app. \ref{app:integrals} we list some integrals that to the best of our knowledge are unknown, and in app. \ref{app:hypergeometric} some further expressions involving hypergeometric functions. In app. \ref{app:w3} we give explicit expressions for all GPLs at weight 3 and some further relations.


\section{Generalized polylogarithms}
\label{sec:definitionGPLs}

GPLs \cite{Poincare:1883, Goncharov:2011ar, Goncharov:2010jf} are defined recursively as
\begin{align}
G(a_1,\ldots,a_n;x) &= \int_0^x \! \frac{\id z}{z-a_1} G(a_2,\ldots,a_n;z)\,,
\label{eq:gdef}
\end{align}
with
\begin{align}
G(\underbrace{0,\dots,0}_{n};x) \,\equiv\,G(\bar{0}_n;x) \,=\, \frac{\log^n(x)}{n!} \;\;\;\;\;\;\;\;\;\; \tekst{and} \;\;\;\;\;\;\;\;\;\; G(;x) \,=\, 1\,,
\label{eq:gdefzeroes}
\end{align}
and with the integration path being a straight line from 0 to the generally complex $x$.

GPLs satisfy a large number of relations between themselves. 
Primary is the rescaling relation
\begin{align}
G(a_1,\ldots,a_n;x) &= G(z a_1,\ldots, za_n; zx)\,,
\label{eq:rescaling}
\end{align}
where $z$ is a general complex number different from zero. The relation is valid whenever $a_n \neq 0$. From eq. \eqref{eq:rescaling} we see that the full set of arguments of a general $G(a_1,\ldots,a_n;x)$ is redundant, as one, conventionally $x$, may always be put to one by setting $z=x^{-1}$. For the case of complex $x$, this $x \rightarrow 1$ rescaling may be used to ensure that the naive integration path of eq. \eqref{eq:gdef} is the correct one.

Additionally there are the shuffle rules \cite{Borwein:1999aa}
\begin{align}
G(a_1,\ldots,a_m;x) G(b_1, \ldots, b_n; x) &= \sum_{c \in a \sha b} G(c_1, \ldots, c_{m+n}; x)\,,
\label{eq:shuffle}
\end{align}
where $a \sha b$ denotes the \emph{shuffle product} of the lists $a$ and $b$, which is defined as the set of all lists containing exactly the elements of $a$ and $b$, for which the ordering of the elements of $a$ and $b$ are preserved.

Note that while the rescaling property of eq. \eqref{eq:rescaling} is not directly valid for $a_n \neq 0$, we may apply it if we first shuffle the zero away using eq. \eqref{eq:shuffle}. Only if all the $a_i$ are zero we cannot rescale $x\rightarrow1$, but in this case we can use the definition of eq. \eqref{eq:gdefzeroes}. In practice we can therefore study $G(a_1,\dots,a_n;1)$ without any loss of generality.

GPLs are equivalent to another class of functions, the multiple polylogarithms (MPLs), which are defined\footnote{We note that the literature is split rather evenly on how to define MPLs. We are using the same definition as e.g. \cite{Vollinga:2004sn, Hoffman:2003ma}, while \cite{Duhr:2012po, Duhr:2012fh} use the opposite definition in which the ordering of the indices is reversed, i.e. $\widehat{\Li}_{m_1,\ldots,m_n}(x_1,\ldots,x_n) = \!\! \sum_{0<k_1<\cdots<k_n}^{\infty} \frac{x_1^{k_1}}{k_1^{m_1}} \cdots \frac{x_n^{k_n}}{k_n^{m_n}}$.} by the sum
\begin{align}
\Li_{m_1,\ldots,m_n}(x_1,\ldots,x_n) &= \!\! \sum_{k_1>\cdots>k_n>0}^{\infty} \frac{x_1^{k_1}}{k_1^{m_1}} \cdots \frac{x_n^{k_n}}{k_n^{m_n}}.
\label{eq:mpldef}
\end{align}
The relation between the two classes is
\begin{align}
\Li_{m_1,\ldots,m_n}(x_1,\ldots,x_n) &= (-1)^n G \Big( \bar{0}_{m_1-1},\tfrac{1}{x_1},\bar{0}_{m_2-1},\tfrac{1}{x_1 x_2},\ldots,\bar{0}_{m_n-1},\tfrac{1}{\prod_{i=1}^n x_i};1 \Big)
\label{eq:mplrelation}
\end{align}
or correspondingly
\begin{align}
G \Big( \bar{0}_{m_1},a_1,\bar{0}_{m_2},a_2,\ldots,\bar{0}_{m_n},a_n; x \Big) &= (-1)^n \Li_{m_1+1,\ldots,m_n+1} \left( \tfrac{x}{a_1}, \tfrac{a_1}{a_2}, \ldots, \tfrac{a_{n-1}}{a_n} \right).
\label{eq:gplrelation}
\end{align}

The MPLs obey a class of relations denoted as the stuffle\footnote{The word 'stuffle' may be obtained by combining the word 'shuffle' with the word 'stuck' which is what the operator $\kur{M}(x)$ of eq. \eqref{eq:stufflepr} makes the elements of the list $x$.} rules \cite{Borwein:1999aa, Duhr:2014woa}. Just as the shuffle rules were based on the shuffle product, the stuffle rules are based on the \emph{stuffle product} defined as
\begin{align}
a*_{\circ}b \, &= \, \bigcup_{j=0} \kur{M}_{a,b,\circ}^j \big( a \sha b \big)\,,
\label{eq:stufflepr}
\end{align}
where $a$ and $b$ are lists, $\circ$ is an operator, and where the operator $\kur{M}_{a,b,\circ}(x)$ acting on a list $x$, gives the set of all list which may be obtained by taking two adjacent elements of $x$ and turning them into one element equaling the original two joined by the operator $\circ$, under the condition that one of the two elements come from $a$, and the other from $b$. We see that the maximal value taken by $j$ equals the length of the shortest of the lists $a$ and $b$. Additionally we define $\kur{M}$ of a set of lists (the way it is applied in eq. \eqref{eq:stufflepr}) to be the union of the results of applying $\kur{M}$ to the individual members of the set, i.e.
\begin{align}
\kur{M} \big( \{ x_1, \ldots, x_n \} \big) &\define \bigcup_{i=1}^{n} \kur{M} (x_i).
\end{align}
We note that the operator $*$ should be applied at the purely symbolic level, which means that even if some members of $a$ and/or $b$ are numerically identical they should still be treated as different by the operator.

With these definitions and considerations in place, the stuffle rules are given as
\begin{align}
\Li_{m_1,\ldots,m_a}(x_1,\ldots,x_a) \Li_{n_1,\ldots,n_b}(y_1,\ldots,y_b) &= \sum_{i} \Li_{u_i} (z_i)\,,
\label{eq:stuffle}
\end{align}
where the sets of lists
\begin{align}
u \,=\, m *_+ n \;\;\;\;\;\;\;\; \tekst{and} \;\;\;\;\;\;\;\; z \,=\, x *_{\times} y
\end{align}
are assigned the same ordering.

An example of the use of the stuffle rules is the relation
\begin{eqnarray}
&& \Li_{m_1,m_2,m_3}(x_1,x_2,x_3) \,\Li_{m_4}(x_4) \;=\; \text{Li}_{{m_1},{m_2},{m_3},{m_4}}\left(x_1,x_2,x_3,x_4\right) \nn \\
&& \;\; +\,\text{Li}_{{m_1},{m_2},{m_4},{m_3}}\left(x_1,x_2,x_4,x_3\right) + \text{Li}_{{m_1},{m_4},{m_2},{m_3}}\left(x_1,x_4,x_2,x_3\right) \\
&& \;\; +\,\text{Li}_{{m_4},{m_1},{m_2},{m_3}}\left(x_4,x_1,x_2,x_3\right) + \text{Li}_{{m_1},{m_2},{m_3}+{m_4}}\left(x_1,x_2,x_3\cdot x_4\right) \nonumber\\
&& \;\; +\,\text{Li}_{{m_1},{m_2}+{m_4},{m_3}}\left(x_1,x_2\cdot x_4,x_3\right) + \text{Li}_{{m_1}+{m_4},{m_2},{m_3}}\left(x_1\cdot x_4,x_2,x_3\right)\,,\nonumber
\end{eqnarray}
where the first four terms correspond to $j=0$ in eq. \eqref{eq:stufflepr}, while the last three correspond to $j=1$.
Note that the stuffle algebra and the shuffle algebra are independent structures \cite{Borwein:1999aa,Ablinger:2013cf}, giving complementary relations among the MPLs and consequently the GPLs.

One additional identity between the GPLs is the H{\"o}lder relation \cite{Borwein:1999aa}
\begin{align}
G(a_1,\ldots,a_n;1) &= \sum_{j=0}^n (-1)^j G(1-a_j,1-a_{j-1},\ldots,1-a_1;1-q) G(a_{j+1},\ldots,a_n;q)\,,
\label{eq:holder}
\end{align}
which holds when $a_1 \neq 1$ and $a_n \neq 0$, and where $q$ may take values in a subset of $\mathbb{C}$ which includes the real numbers.

GPLs can be assigned the property of {\it weight} corresponding to the number of logarithmic integrations \cite{Duhr:2014woa} in \eqref{eq:gdef}, such that $G(a_1,\ldots,a_n;x)$ has weight $n$. This implies that $\log(x)$ has weight $1$, $\Li_n(x)$ has weight $n$, and $\Li_{m_1,\ldots,m_n}(x_1,\ldots,x_n)$ has weight $\sum_{i=1}^n m_i$. Additionally a product of two functions with respective weights $m$ and $n$ is assigned weight $m+n$. We see that most of the equations in this paper conserve this quantity explicitly.

Some useful relations can be obtained by integration by parts
\begin{align}
G(z_1,z_2,\ldots,z_n;1)&=G(z_1;1)G(z_2,z_3,\ldots,z_n;1)-\!\int_0^1\! \frac{G(z_1;t)G(z_3,\ldots,z_n;t)}{t-z_2} \id t
\label{eq:GIBP}\\
=G(z_1;1)G(z_2,\ldots&,z_n;1)\!-\!G(z_2,z_1;1)G(z_3,\ldots,z_n;1)\!+\!\ldots\!-\! (\!-\!1)^{n}G(z_n,\dots,z_1;1). \nn
\end{align}

Note from the last expression that the sum $G(z_1,z_2,\ldots,z_n;1)+(-1)^{n}G(z_n,\dots,z_1;1)$ can be expressed as a combination of lower weight GPLs.
Furthermore, following eq. (\ref{eq:gdef}) we may evaluate GPLs of weight $n$ by an $n$-dimensional numerical integration; on the other hand by iterating the expression of eq. (\ref{eq:GIBP}) (and recalling that $G(z;t)=\log(1-t/z)$) we may reduce the total number of dimensions of the integration from $n$ to $n/2$. A possible numerical implementation of a weight $n$ GPL is by evaluating the weight $n-1$ GPL that appears on the r.h.s. of eq. \eqref{eq:gdef} by its series representation as in eqs. (\ref{eq:mpldef}, \ref{eq:gplrelation}) and then numerically performing the last single integration in eq. \eqref{eq:gdef}. By the use of eq. (\ref{eq:GIBP}) the GPLs only need to be evaluated up to weight $n-2$ instead, and then integrated over one dimension in order to evaluate a GPL of weight $n$.

A further relation can be obtained from eq. (\ref{eq:GIBP}) in case of one or more zero letters
\begin{align}
G(\bar 0_m,z_1,z_2,\ldots,z_n;1)&=\frac{(-1)^{m}}{m!}\int_0^1 \frac{\log(t)^m}{(t-z_1)}G(z_2\ldots,z_n;t) \id t.
\label{eq:GIBP4}
\end{align}

As can be seen from eq. \eqref{eq:gdef}, the GPLs are in general not well defined whenever any of the letters $a_1,\ldots,a_n$ lie exactly along the integration path that is the straight line in complex space connecting the origin and the argument $x$, i.e. if $a_i/x\in(0,1)$. There is a discontinuity whenever $a_i$ crosses the straight line connecting the origin and the (fixed) argument $x$ and these lines also define exactly all the branch cuts of the GPLs. Therefore, whenever $a_i/x\in(0,1)$ an infinitesimal perturbation in $x$ (or equivalently in the letters because of eq. \eqref{eq:rescaling}) is required such as to make the GPL well defined. An example of such a perturbation is to multiply the argument by a factor $1\pm i\eps$ where afterwards the limit $\eps\rightarrow 0$ is taken. The usual adopted convention is to take $1-i\eps$ \cite{british2003c,Vollinga:2004sn}.

Up to weight three, it is known \cite{lewin1981, Duhr:2012po} that all GPLs can be expressed as combinations of the functions $\log(x)$, $\Li_2(x)$, and $\Li_3(x)$. In \cite{Duhr:2012po} a minimal basis of functions at higher weights were proposed, and specifically it was proposed that at weight four the additional basis functions $\Li_4(x)$ and $\Litt(x,y)$ are required, where the latter, in accordance with eq. \eqref{eq:mplrelation}, may be expressed as\footnote{As is the case for the general MPL (see the note to eq. \eqref{eq:mpldef}), the literature is split on how to define $\Litt$. The alternative definition has the two arguments exchanged such that $\widehat{\Li}_{2,2}(x,y) = G \big( 0, \tfrac{1}{y}, 0, \tfrac{1}{xy} ; 1 \big)$. }
\begin{align}
\Litt(x,y) &= G \big( 0, \tfrac{1}{x}, 0, \tfrac{1}{xy} ; 1 \big) .
\label{eq:li22asg}
\end{align}
Notice that for instance $\Li_{3,1}$ (or $\Li_{1,3}$) may be chosen as a basis functions instead of $\Li_{2,2}$ since they are related as given in eq. \eqref{eq:Li31repl}. In the following two sections, \ref{sec:newrelationsGPLs} and \ref{sec:gpl}, we will show that this conjecture is correct and also present some explicit expressions for GPLs up to weight four.

The relations presented in this section are not the only ones satisfied by the GPLs. Numerous other structural relations exist which are discussed in detail elsewhere, see for instance \cite{Borwein:1999aa, Vollinga:2004sn, Duhr:2012po,Ablinger:2013cf, Panzer:2014caa}.


\section{Some reduction relations for GPLs}
\label{sec:newrelationsGPLs}

In order to reduce the GPLs we may use some powerful relations that the GPLs satisfy. We have already discussed the rescaling \eqref{eq:rescaling}, shuffle \eqref{eq:shuffle} and stuffle \eqref{eq:stuffle} identities in the previous section. Based on these identities alone, all GPLs can be expressed in terms of a smaller set of functions and we choose the following basis
\begin{eqnarray}
&\text{weight 1}:& G(a;1) \nonumber\\
&\text{weight 2}:& G(a,b;1),\, G(0,a;1) \nonumber\\
&\text{weight 3}:& G(a,b,c;1),\, G(0,a,b;1),\, G(0,0,a;1) \nonumber\\
&\text{weight 4}:& G(a,b,c,d;1),\, G(0,a,b,c;1),\, G(0,a,0,b;1),\, G(0,0,0,a;1)\,,
\end{eqnarray}
where $a,b,c,d$ are different non-zero complex numbers.

\begin{figure}[t!]
\centering
\includegraphics[width=0.35\textwidth]{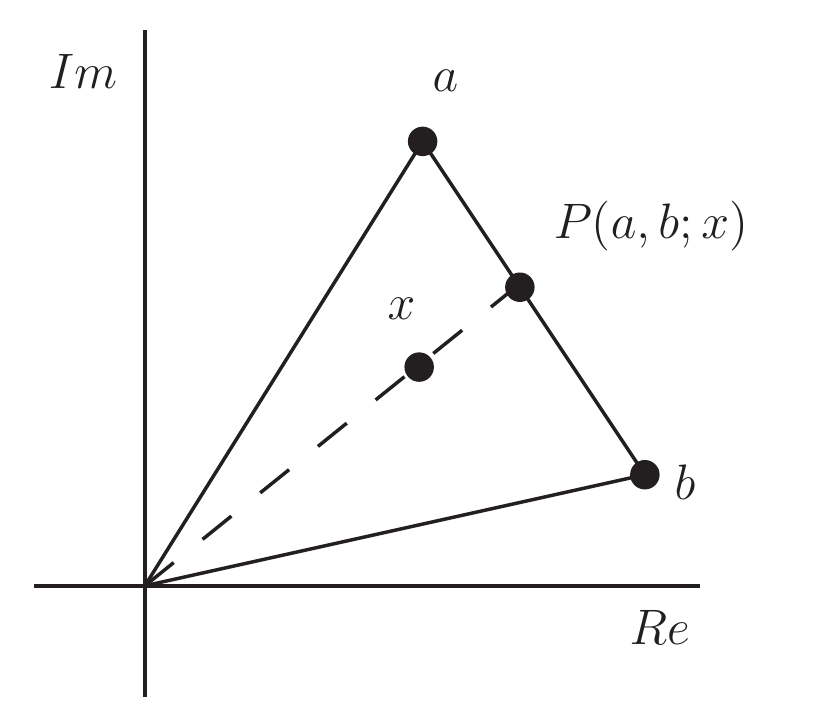}
\caption{This figure shows the triangle that is spanned by the points $0$, $a$ and $b$ in the complex plane. The function $\text{T}$ as given by eq. \eqref{eq:triangleT} evaluates to one whenever the point $x$ is inside the triangle, and to zero otherwise. $\text{P}(a,b;x)$ denotes the point where the radial line going through $x$ and the line between $a$ and $b$ cross.}
\label{fig:triangleT}
\end{figure}

For a further reduction of the above basis integrals, we can use relations found by recursively expressing GPLs with argument $1-x$ in terms of GPLs with argument $x$. These and other similar transformation rules were discussed in ref.~\cite{Vollinga:2004sn}, albeit not in the context of GPL reduction. At weights one and two the relations take the following form
\begin{gather}
G(1;1-x)=G(0;x)\,, \label{eq:1mxw2}\\
G(a,1;1-x)=G(1-a,0;x)-G(1-a,0;1)-2\pi i\, \text{sgn}(\tekst{Im}(a))\, \text{T}(1,x;1-a)\, G(0;1-a). \nonumber
\end{gather}
The function $\text{T}(a,b;x)$ equals one whenever the point $x$ lies inside the triangle spanned by the three points $0,a$ and $b$ in the complex plane and zero otherwise (see fig. \ref{fig:triangleT}). It may be expressed as
\begin{gather}
\text{T}(a,b;x) \, \define \, \theta \! \left(\tfrac{\tekst{Im}(\bar{x}a)}{\tekst{Im}(\bar{x}(a-b))}\right) \theta \! \left(1-\tfrac{\tekst{Im}(\bar{x}a)}{\tekst{Im}(\bar{x}(a-b))}\right) \theta \! \left(\tfrac{\tekst{Im}(\bar{a}b)}{\tekst{Im}(\bar{x}(a-b))}-1\right)\,, \label{eq:triangleT}
\end{gather}
where $\bar{x}$ denotes the complex conjugate of $x$. At weight three the corresponding relation is
\begin{gather}
G(a,b,1;1-x)=G(1-b,0;1) \Big(G(1-a;1)-G(1-a;x)\Big)+G(1-a,1-b,0;x) \nonumber\\ 
-G(1-a,1-b,0;1)-2 \pi  i\, \text{sgn}(\tekst{Im}(a))\, \text{T}(1,x;1-a)\, \Big(G(1-b,0;1-a)-G(1-b,0;1)\Big) \nonumber\\
-2 \pi  i\, \text{sgn}(\tekst{Im}(b))\, \text{T}(1,x;1-b)\, G(0;1-b) \Big(G(1-a;x)-G(1-a;1-b) \nonumber\\
-2 \pi  i\, \text{sgn}(\tekst{Im}(a))\, \text{T}(\text{P}(1,x;1-b),x;1-a)\Big), \label{eq:1mxw3} 
\end{gather}
with
\begin{align}
\text{P}(a,b;x) &\define \frac{\tekst{Im}(\bar{a}b)}{\tekst{Im}(\bar{x}(a-b))}x\,,
\end{align}
see fig. \ref{fig:triangleT} for the geometric meaning of the point $\text{P}(a,b;x)$.

Finally at weight four the relation reads
\begin{gather}
G(a,b,c,1;1-x)=G(1-a,1-b,1-c,0;x)-G(1-a,1-b,1-c,0;1) \nonumber\\
-G(1-a;x) G(1-b,1-c,0;1)+G(1-a;1) G(1-b,1-c,0;1) \nonumber\\
+G(1-c,0;1) \Big(G(1-b;1) G(1-a;x)-G(1-a,1-b;x)-G(1-b,1-a;1)\Big) \nonumber\\
-2\pi i\, \text{sgn}(\tekst{Im}(a)) \, \text{T}(1,x;1-a) \bigg(\! G(1-b,1-c,0;1-a)\!-\!G(1-b,1-c,0;1)\!+\!G(1-c,0;1) \nonumber\\
\times\Big(G(1-b;1)-G(1-b;1-a)\Big)\!\bigg)-2 \pi i\, \text{sgn}(\tekst{Im}(b))\,\text{T}(1,x;1-b) \Big(G(1-c,0;1-b) \nonumber\\
-G(1-c,0;1)\!\Big)\! \Big(\! \! -\! G(1-a;1-b)\! +\! G(1-a;x)\!-\! 2 \pi i\, \text{sgn}(\tekst{Im}(a)) \text{T} \big( \text{P}(1,x;1-b),x;1-a \big)\! \Big) \nonumber\\
-2 \pi i\,\text{sgn}(\tekst{Im}(c)) \text{T}(1,x;1-c)  G(1-c;0) \bigg(G(1-a,1-b;x)+G(1-b,1-a;1-c) \nonumber\\
-G(1-a;x) G(1-b;1-c)-2 \pi i\, \text{sgn}(\tekst{Im}(a)) \text{T} \big( \text{P}(1,x;1-c),x;1-a \big) \Big(G(1-b;1-a) \nonumber\\
-G(1-b;1-c)\Big)-2 \pi i\, \text{sgn}(\tekst{Im}(b)) \text{T} \big( \text{P}(1,x;1-c),x;1-b \big) \Big(G(1-a;x)-G(1-a;1-b) \nonumber\\
-2 \pi i\, \text{sgn}(\tekst{Im}(a)) \text{T} \big( \text{P}(1,x;1-b),x;1-a \big)\Big)\bigg). \label{eq:1mxw4}
\end{gather}
In the relations \eqref{eq:1mxw2}, \eqref{eq:1mxw3} and \eqref{eq:1mxw4} above, the letters $a,b,c$ are assumed to be different from each other and non-zero. The relations follow a pattern and can be extended to these cases and also to higher weights. The crucial point here is that the above formulae are exact for {\it any} complex values of the parameters\footnote{Whenever $\tekst{Im}(\bar{x}(a-b))=0$, one may safely set $\text{T}=0$.}. This is achieved by the introduction of the ``triangle'' function $\text{T}$ in \eqref{eq:triangleT}, which appears naturally when performing the integrals recursively while taking into account the pole of the integrand. In particular, the functions $\text{T}$ appearing in the expressions above cancel exactly the discontinuities~\cite{Duhr:2012fh} along {\it spurious} branch cuts, in the complex plane of the letters $a,b,c$ for any fixed $x$, that are introduced by the various GPLs in the expression. At the same time they also correctly reproduce the discontinuities along the true branch cuts of the corresponding GPL. The simplest case where this can be seen is the weight two formula \eqref{eq:1mxw2}: the branch cuts in the complex plane of $a$ along the two lines connecting the pairs of points $(1,1-x)$ and $(0,1)$ respectively are removed by the last term $-2 \pi  i\, \text{sgn}(\tekst{Im}(a))\,\text{T}(1,x;1-a)\,G(0;1-a)$. Simultaneously, this same term correctly reproduces\footnote{In order to see this, note that the three lines connecting the pairs of points $(1,1-x)$, $(0,1)$ and $(0,1-x)$ in the complex space of $a$ corresponds to three lines connecting the pairs of points $(0,x)$, $(0,1)$ and $(1,x)$ in the complex space of $1-a$ which equals exactly the three sides of the triangle defined by $\text{T}(1,x;1-a)$ in fig. \ref{fig:triangleT}. Using the fact that the discontinuity of the functions $\log(x)$ and $\Li_2(x)$ along their branch cuts equal $2\pi i$ and $2\pi i \log(x)$ respectively, one may see that the discontinuity of $G(1-a,0;x)=G(1-a;x)G(0;x)-G(0,1-a;x)$ and $G(1-a,0;1)=-G(0,1-a;1)$ along the lines connecting the points $(1,1-x)$ and $(0,1)$ respectively is exactly $2\pi i\,G(0;1-a)$. It can similarly be shown that $2\pi i \,G(0;1-a)$ is the correct discontinuity of $G(a,1;1-x)$ along the true branch cut.} the discontinuity along the branch cut connecting the true branch points $0$ and $1-x$.

Whenever any of the letters $a,b,c$ lie exactly on the lines defining the triangle function $\text{T}$, one encounters an undefined $\theta(0)$ and an infinitesimal perturbation of $x$ is required to remove the letter off of the line such that the appearance of $\theta(0)$ is avoided\footnote{For the $\text{sgn}$ factors in the expression a zero as argument is not an issue, since whenever $\text{sgn}(0)$ appears it will multiply a vanishing $\text{T}$ function.}. If the letters lie along a {\it spurious} branch cut, the result will be {\it independent} of the exact value of the perturbation in $x$, as long as it is small enough and does not result in the crossing of a true branch cut by the letters. However if any of the letters lie along a true branch cut of the GPL, i.e. if for example $a/(1-x)\in (0,1)$, a perturbation in $x$ is required to make the GPL well defined to begin with (we refer to the discussion below eq. \eqref{eq:GIBP4}), which will then make sure that again a $\theta(0)$ in $\text{T}$ is avoided. These considerations are for example already required for the stuffle identities \eqref{eq:stuffle} and similar issues have also been discussed about in the context of GPL identities~\cite{Duhr:2012po} and Feynman integrals~\cite{Bonciani:2015eua}. Lastly it is important to note that {\it all} the triangle functions $\text{T}$ in the relations \eqref{eq:1mxw2}, \eqref{eq:1mxw3} and \eqref{eq:1mxw4} vanish whenever the absolute values of the letters $a,b,c$ are larger than one and the absolute value of the argument $1-x$.

The letters appearing in the above $1-x$ transformation rules are all of the form one minus the letter in the original GPL. In particular the GPLs of highest weight that appear on the right hand sides of the relations have one letter equal to zero. Therefore if we start with any GPL up to weight four and then rescale for example its rightmost letter to one, assuming it is non-zero, and afterwards apply \eqref{eq:1mxw2}, \eqref{eq:1mxw3} and/or \eqref{eq:1mxw4}, one may express {\it any} GPL of weight $n\leq 4$ in terms of a sum of weight $n$ GPLs where at least one letter is set to zero plus a combination of lower weight GPLs. In general when the GPL has $n_0$ letters that are the same, we may rescale all of them to one and apply similar relations as above to express it in terms of weight $n$ GPLs with all least $n_0$ zero letters plus a combination of lower weight GPLs. In this way the basis integrals are further reduced to
\begin{eqnarray}
&\text{weight 1}:& G(a;1) \nonumber\\
&\text{weight 2}:& G(0,a;1) \nonumber\\
&\text{weight 3}:& G(0,a,b;1), \, G(0,0,a;1) \nonumber\\
&\text{weight 4}:& G(0,a,b,c;1), \, G(0,a,0,b;1), \, G(0,0,0,a;1). \label{eq:prelbasis}
\end{eqnarray}

Before continuing we will mention one alternative to the $1-x$ identities. That is applying the H\"older convolution \eqref{eq:holder}, such that any GPL of arbitrary weight $n$ may be expressed in terms of GPLs with weight $n$ with at least one letter equal to zero, plus a combination of lower weight GPLs
\begin{gather}
G\left(z_1,\ldots,z_{n-1},1; x \right) =  G\left(z_1,\ldots,z_{n-1},1; 1 \right)- (-1)^nG\left(0,1-z_{n-1},\ldots,1-z_1; 1 - x \right) \nonumber\\
 -\sum\limits_{i=1}^{n-1} \left(-1\right)^i G\left(1-z_i, 1-z_{i-1},...,1-z_1; 1 - x \right)G\left( z_{i+1},..., z_k; x \right) \nonumber\\
= (-1)^n (G\left(0,1-z_{n-1},\ldots,1-z_1; 1 \right)- G\left(0,1-z_{n-1},\ldots,1-z_1; 1 - x \right)) \nonumber\\
  -\sum\limits_{i=1}^{n-1} \left(-1\right)^i G\left(1-z_i, 1-z_{i-1},...,1-z_1; 1 - x \right)G\left( z_{i+1},..., z_k; x \right).
\end{gather}
We have applied \eqref{eq:holder} twice on the term $G\left(z_1,\ldots,z_{n-1},1; 1 \right)$ in the above equation, namely once for the first equation with $q=x$ and once again with $q=0$ resulting in the second equation, both times assuming $z_1\neq 1$. Yet for our reductions we have not used the H\"older convolution but instead the $1-x$ identities in \eqref{eq:1mxw2}, \eqref{eq:1mxw3} and \eqref{eq:1mxw4} as previously mentioned. 

In the next section we give expressions for $G(0,a,b;1)$ and $G(0,a,b,c;1)$ that can be reduced to classical polylogarithms and $\Li_{2,2}$. Therefore the final set of basis functions up to weight four will be shown explicitly to be the logarithm, the classical polylogarithms $\Li_2$, $\Li_3$, and $\Li_4$, and the function $\Li_{2,2}$, which can itself {\it not} be further reduced to polylogarithms and which therefore may be considered as a candidate for a standard function.


\section{Expressions for GPLs}
\label{sec:gpl}

In this section we list expressions for some GPLs of weight $\leq 4$, that are valid for all values of the complex parameters and depend only on rational combinations of the letters. 
At weight one, two expressions are enough to describe the entire complex domain
\begin{align}
G(0;x) &= \log(x) & G(a;x) &= \log \left( 1 - \tfrac{x}{a} \right)\quad \mbox{for } a \neq 0\,.
\end{align}

At weight two we likewise have to have separate expressions depending on whether the letters are zero or non-zero, and also depending on whether or not they are equal
\begin{align}
G(0,0;x) &= \tfrac{1}{2} \log^2(x) & G(0,a;x) &= - \Li_2 \! \left(\tfrac{x}{a}\right) \nn \\
G(a,0;x) &= \log(x) \log \! \left( 1 - \tfrac{x}{a} \right) + \Li_2 \! \left(\tfrac{x}{a}\right) & G(a,a;x) &= \tfrac{1}{2} \log^2 \! \left( 1 - \tfrac{x}{a} \right) 
\end{align}
\vspace{-0.8cm}
\begin{align}
G(a,b;x) &= \Bigg\{ \begin{array}{ll} \Li_2 \! \left( \frac{b - x}{b - a} \right) - \Li_2 \! \left( \frac{b}{b - a} \right) + \log \! \left( 1 - \frac{x}{b} \right) \log \! \left( \frac{x - a}{b - a} \right) & \;\; \tekst{when} \;\;\; \left|\tekst{Im}(\frac{a}{x})\right|>\left|\tekst{Im}(\frac{b}{x})\right| \\ 
\Li_2 \! \left( \frac{a}{a - b} \right) - \Li_2 \! \left( \frac{a - x}{a - b} \right) + \log \! \left( 1 - \frac{x}{a} \right) \log \! \left( \frac{b - a}{b} \right) & \;\; \tekst{otherwise}. \end{array} \nonumber
\end{align}
The latter of the above expressions is our first example in which a single expression is insufficient to cover all complex values of the function arguments. However, if we admit Heaviside $\theta$ functions in our expressions we may use eq. \eqref{eq:1mxw2} to write the latter expression as
\begin{eqnarray}
G(a,b;x)& = & \Li_2 \! \left( \tfrac{b - x}{b - a} \right) - \Li_2 \! \left( \tfrac{b}{b - a} \right) + \log \! \left( 1 - \tfrac{x}{b} \right) \log \! \left( \tfrac{x - a}{b - a} \right) \nonumber\\
& & +2\pi i\,\text{sgn}\left( \tekst{Im} \! \left(\tfrac{b}{x}\right)\! \right) \log\left(1-\tfrac{a}{b}\right) \text{T}\left(1,1-\tfrac{x}{b};1-\tfrac{a}{b}\right), \label{eq:Gw2red}
\end{eqnarray}
where the function $\text{T}$ is given in the previous section.

This approach of writing the GPLs as single expressions with possible $\theta$ functions is so general that it extends to weight four as we will show explicitly in the following. For the remainder of this section we will without loss of generality rescale the argument $x$ to one. At weight three we may apply eq. \eqref{eq:1mxw3} to express a GPL in terms of weight three GPLs with at least one letter being zero plus a combination of lower weight GPLs. The remaining function which is required according to eq. \eqref{eq:prelbasis} is therefore
\begin{eqnarray}
G(0,a,b;1)&=&-\Li_3 \! \left(\tfrac{a-a b}{a-b}\right)-\Li_3 \! \left(-\tfrac{b}{a-b}\right)+\Li_3\! \left(\tfrac{b-1}{b-a}\right)+\Li_3 \! \left(\tfrac{1}{a}\right)+\Li_3(1-b) \nonumber\\
&&+\log \left(\tfrac{b-1}{b}\right) \! \left( \Li_2 \! \left(\tfrac{a-a b}{a-b}\right)-\Li_2 \! \left(\tfrac{b-1}{b-a}\right)-\Li_2(1-b) \right)\! -\tfrac{1}{6} \log ^3\! \left(\tfrac{a b}{a-b}\right) \nonumber\\
&&+\tfrac{1}{2}\log^2\!\left(\tfrac{b-1}{b}\right) \! \left( \log \! \left(\tfrac{(a-1) b}{a-b}\right)-\log (b)-\log \! \left(\tfrac{a-1}{a-b}\right)\! \right)\! -\tfrac{1}{6} \pi ^2 \log \! \left(\tfrac{a b}{a-b}\right) \nonumber\\
&&+\tfrac{\log ^3(b)}{6}+\tfrac{1}{6} \pi ^2 \log (b)+i \pi  \log ^2 \! \left(\tfrac{b-a}{a b}\right) \text{sgn}(\tekst{Im}(b)) \, \mathcal{H}_1(a,b) \nonumber\\
&&+i \pi  \log ^2 \! \left(\tfrac{b-a}{b}\right) \text{T} \! \left(1,1-\tfrac{1}{b},1-\tfrac{a}{b}\right) \text{sgn} \! \left(\tekst{Im}\left(\tfrac{a}{b}\right)\right). \label{eq:G0ab}
\end{eqnarray}
The above expression may be obtained by extending the derivations and results in section 8.4 of \cite{lewin1981} to the whole complex plane. It is valid whenever $a$ and $b$ are different and non-zero. The function $\mathcal{H}_1$ that appears in eq. \eqref{eq:G0ab} above is defined as\footnote{Whenever $\tekst{Im}(\bar{a}b)=0$, one may safely set $\mathcal{H}_1=0$.}
\begin{gather}
\mathcal{H}_1(a,b) \define \theta \! \left(\min \left(1,\tfrac{\left|a\right|^2 \tekst{Im}(b)}{\tekst{Im}(\bar{a}b)}\right)-\text{r}(a,b)\right)\theta \! \left(\text{r}(a,b)\right)\,, \nonumber\\
\text{r}(a,b) \define \tfrac{\left|a\right|^2 \tekst{Im}(b)-\left|b\right|^2 \tekst{Im}(a)}{\tekst{Im}(\bar{a}b)}\,.
\end{gather}
The above expressions explicitly reproduce the known result \cite{lewin1981, Duhr:2012po} that GPLs up to weight three can be expressed in terms of $\Li_3,\Li_2$ and logarithms for any complex values of the parameters. The explicit results for all GPLs at weight 3 are given in app. \ref{app:w3}.

According to the list \eqref{eq:prelbasis}, at weight four we need to compute the function $G(0,a,b,c;1)$, which is related by shuffle identities to $G(a,b,c,0;1)$. Assume henceforth that the arbitrary complex letters $a,b,c$ are different and non-zero. In order to compute $G(0,a,b,c;1)$, $G(a,0,b,c;1)$, $G(a,b,0,c;1)$ and $G(a,b,c,0;1)$ we consider directly their definition in eq. \eqref{eq:gdef} and plug in the analytic expressions of the weight three GPLs found above. By subtracting contributions to the above four GPLs that do not integrate directly to $\Li_{2,2}$ or $\Li_{n\leq4}$ by standard integration techniques, and relating the remainders denoted below as $\tilde{G}$, by stuffle identities it follows that $G(a,b,c,0;1)$ can be expressed as
\begin{gather}
G(a,b,c,0;1)=-G\left(0,a,0,\tfrac{c}{b};1\right)+\tfrac{1}{2} G\left(0,a,0,\tfrac{a c}{b};1\right)-G\left( 0,a,0,\tfrac{b}{c};1 \right)\! -\! G\left(a,0,0,\tfrac{c}{b};1\right)\nonumber\\
+\tfrac{1}{2} G\left(0,a,0,\tfrac{a b}{c};1\right)+\tfrac{1}{2} G\left(0,\tfrac{c}{b},0,\tfrac{c}{a b};1\right)-\tfrac{1}{2} G\left(0,\tfrac{a c}{b},0,a;1\right)-\tfrac{1}{2} G\left(0,\tfrac{c}{a b},0,\tfrac{c}{b};1\right) \nonumber\\
+\tfrac{1}{2} G\left(0,\tfrac{b}{c},0,\tfrac{b}{a c};1\right)-\tfrac{1}{2} G\left(0,\tfrac{a b}{c},0,a;1\right)-\tfrac{1}{2} G\left(0,\tfrac{b}{a c},0,\tfrac{b}{c};1\right)-G\left(0,0,0,\tfrac{a}{b};1\right) \nonumber\\
+\tfrac{1}{2} G\left(0,\tfrac{b}{a},0,b;1\right)-\tfrac{1}{4} G\left(0,a,0,\tfrac{a}{b};1\right)+\tfrac{3}{4} G(0,a,0,b;1)-\tfrac{1}{2} G\left(0,\tfrac{1}{b},0,\tfrac{a}{b};1\right) \nonumber\\
-\tfrac{1}{4} G\left(0,\tfrac{a}{b},0,a;1\right)+\tfrac{1}{2} G\left(0,\tfrac{a}{b},0,\tfrac{1}{b};1\right)-\tfrac{1}{2} G\left(0,b,0,\tfrac{b}{a};1\right)-\tfrac{1}{4} G(0,b,0,a;1) \nonumber\\
+\tfrac{1}{2} G\left(a,0,\tfrac{a}{b},0;1\right)+\tfrac{1}{2} G\left(a,\tfrac{a}{b},0,0;1\right)-G\left(0,0,0,\tfrac{a}{c};1\right)+\tfrac{1}{2} G\left(0,\tfrac{c}{a},0,c;1\right) \nonumber\\
-\tfrac{1}{4} G\left(0,a,0,\tfrac{a}{c};1\right)+\tfrac{3}{4} G(0,a,0,c;1)-\tfrac{1}{2} G\left(0,\tfrac{1}{c},0,\tfrac{a}{c};1\right)-\tfrac{1}{4} G\left(0,\tfrac{a}{c},0,a;1\right) \nonumber\\
+\tfrac{1}{2} G\left(0,\tfrac{a}{c},0,\tfrac{1}{c};1\right)-\tfrac{1}{2} G\left(0,c,0,\tfrac{c}{a};1\right)-\tfrac{1}{4} G(0,c,0,a;1)-\tfrac{1}{2} G(a,0,c,0;1) \nonumber\\
-\tfrac{3}{2} G(a,c,0,0;1)-4 G(0,0,0,a;1)-G\left(0,0,0,\tfrac{c}{b};1\right)-G\left(0,0,0,\tfrac{b}{c};1\right) \nonumber\\
+G(0,0,0,b;1)+G(0,0,0,c;1) +\tfrac{1}{2} \left(-\tilde{G}\left(a,0,\tfrac{a}{b},\tfrac{a}{c};1\right)+\tilde{G}\left(a,0,\tfrac{b}{c},b;1\right) \right. \nonumber\\
\left. +\tilde{G}\left(a,0,\tfrac{b}{c},\tfrac{a}{c};1\right)+\tilde{G}(a,0,c,b;1)\right)+\text{S}(a,b,c,0;1)\,.
\label{eq:Gabc0prel}
\end{gather}
The last function $\text{S}(a,b,c,0;1)$ in eq. \eqref{eq:Gabc0prel} is given in app. \ref{app:w3} and is combination of GPLs up to weight 3, while
\begin{eqnarray}
\tilde{G}(a,0,b,c;1)&\define& G(a,0,b,c;1)-\int_0^1\frac{G(c;x)G(0,b;x)}{x-a}dx \nonumber\\
&=& -\int_0^1\frac{G(c,0,b;x)+G(0,c,b;x)}{x-a}dx \label{eq:tildeG1}
\end{eqnarray}
is the remaining contribution from $G(a,0,b,c;1)$ that does integrate directly to $\Li_{2,2}$ and $\Li_{n\leq 4}$ by standard integration techniques. Note that all GPLs shown in eq. \eqref{eq:Gabc0prel} have at least two zero letters and may therefore be expressed in terms of $\Li_{2,2}$, polylogarithms and logarithms. What is thus left is to express $\tilde{G}(a,0,b,c;1)$ in terms of $\Li_{2,2}$ and polylogarithms. For this purpose the following observation is useful, namely whenever $x\in[0,1]$
\begin{gather}
G\left(0,0,1;\tfrac{b (x-c)}{x (b-c)}\right)=-(G(0,c,b;x)+G(c,0,b;x))+G(c,c,b;x)+G(0,0,b;x) \nonumber\\
-G(0,0,c;x)+G(0,c,c;x)+G(c,0,c;x)-G(c,c,c;x) -\tfrac{1}{6} \log ^3\left(-\tfrac{x (b-c)}{b (x-c)}\right) \label{eq:tildeG2}\\
-\tfrac{1}{6} \pi ^2 \log \left(-\tfrac{x (b-c)}{b (x-c)}\right)-2 \pi i\, \text{sgn}\, (\tekst{Im}(b)) \mathcal{H}_2(b,c) G\left(0,0;\tfrac{x (b-c)}{b (x-c)}\right), \nonumber
\end{gather}
where the function $\mathcal{H}_2$ is defined as
\begin{gather}
\mathcal{H}_2(a,b) \, \define \, \theta \big( -\tekst{Im}(a) \tekst{Im}(b) \big) \theta \big( 1-\text{r}(a,b) \big) \theta \big( \text{r}(a,b) \big)=\lim_{y\rightarrow 0}\text{T}\left(1,-\tfrac{1-a}{a};\tfrac{1-b}{y b}\right). 
\label{eq:H2}
\end{gather}

The first two terms in eq. \eqref{eq:tildeG2} are exactly the integrand in eq. \eqref{eq:tildeG1} and therefore, by direct integration, we finally get
\begin{gather}
\tilde{G}(a,0,b,c;1)=-G(a,c,c,b;1)+G\left(\tfrac{a (b-c)}{b (a-c)},0,0,1;\tfrac{b-c}{b-b c}\right)-G(a,0,0,b;1) \nonumber\\
+G(a,0,0,c;1)-G(a,0,c,c;1)-G(a,c,0,c;1)+G(a,c,c,c;1) -G\left(1-\tfrac{c}{b},0,0,1;\tfrac{b-c}{b-b c}\right) \nonumber\\
+\mathcal{H}_2(b,c) \Bigg\{4 \pi ^2\, \text{sgn}(\tekst{Im}(b))\, \text{sgn}(\tekst{Im}(c))\, \mathcal{H}_2(c,a)\, G\left(0,0;\tfrac{a (b-c)}{b (a-c)}\right) \theta(\text{r}(c,a)-\text{r}(b,c)) \nonumber\\
 -2 \pi i \left\{-\text{sgn}(\tekst{Im}(c)) \left(G\left(\tfrac{a (b-c)}{b (a-c)},0,0;1\right)+G\left(1-\tfrac{c}{b},0,0;1-\tfrac{c^2}{b^2}\right)-G\left(1-\tfrac{c}{b},0,0;1\right) \right. \right. \nonumber\\
\left. \left. -G\left(\tfrac{a (b-c)}{b (a-c)},0,0;1-\tfrac{c^2}{b^2}\right)\right) +\text{sgn}(\tekst{Im}(b)) \left(G\left(\tfrac{a (b-c)}{b (a-c)},0,0;1-\tfrac{c^2}{b^2}\right) \right.\right.  \nonumber\\
 \left. \left. +G\left(1-\tfrac{c}{b},0,0;\tfrac{b-c}{b-b c}\right)-G\left(1-\tfrac{c}{b},0,0;1-\tfrac{c^2}{b^2}\right)-G\left(\tfrac{a (b-c)}{b (a-c)},0,0;\tfrac{b-c}{b-b c}\right)\right)\right\}\Bigg\} \nonumber\\
-2 \pi i\,  \text{sgn}(\tekst{Im}(c)) \mathcal{H}_2(c,a) G\left(0,0,\tfrac{b (a-c)}{a (b-c)};1\right). \label{eq:G0abc}
\end{gather}
All GPLs in eq. \eqref{eq:G0abc} have at least two letters equal to zero or two letters which are the same. By again applying eq. \eqref{eq:1mxw4} the GPLs with two equal letters can expressed in terms of GPLs with two letters equal to zero, i.e. the basis functions $\Li_{2,2},\Li_4,\Li_3,\Li_2$ and logarithms. The complete expression for $G(a,b,c,d;1)$ in terms of the basis functions can now be found by combining\footnote{The GPLs of weights one, two and three that appear in the final result can be expressed in terms of the basis functions by our previous results in this section and the expressions given in app. \ref{app:w3}.} eqs. \eqref{eq:1mxw4}, \eqref{eq:Gabc0prel}, \eqref{eq:G0abc} and \eqref{eq:G3abc0}. We refer to the ancillary files for the explicit expressions for $G(0,a,b,c;1)$, $G(a,b,c,d;1)$ and similarly for the cases where some letters coincide. The functions $\mathcal{H}_1$ and $\mathcal{H}_2$ serve a similar purpose as the ``triangle'' function $\text{T}$ for the $1-x$ identities discussed in the previous section. They cancel discontinuities along spurious branch cuts and contribute to the correct discontinuities along true branch cuts in eqs. \eqref{eq:G0ab} and \eqref{eq:tildeG2} respectively. The way of circumventing the appearance of an undefined $\theta(0)$ by small perturbations of the parameters as discussed below eq. \eqref{eq:1mxw4} applies similarly for the eqs.~\eqref{eq:Gw2red}, \eqref{eq:G0ab}, \eqref{eq:tildeG2} and \eqref{eq:G0abc}.

Lastly, we note that eq. \eqref{eq:tildeG2} may be used to derive an alternative expression for $G(0,a,b;1)$ that contains the function $\mathcal{H}_2$, which is itself related to the triangle function $\text{T}$ (cf. eq. \eqref{eq:H2}), instead of the function $\mathcal{H}_1$ as in eq. \eqref{eq:G0ab}. Therefore all GPLs up to weight four are in principle fully expressible in terms of the basis functions and only the triangle function $\text{T}$. However, in our ancillary files we kept the expression \eqref{eq:G0ab} for $G(0,a,b;1)$ in terms of the function $\mathcal{H}_1$ as it is more compact.

To conclude this section, we have explicitly shown here that the conjecture posed in \cite{Duhr:2012po} is correct, namely that all GPLs up to weight four are expressible in terms of $\Li_{2,2},\Li_4,\Li_3,\Li_2$ and logarithms. Furthermore, in the ancillary files we give explicitly a replacement rule which can be applied to map {\it any} GPL up to weight four to the basis functions for all complex values of the parameters. With these GPLs fully reduced, we will in the remaining sections discuss the algorithms we use to numerically evaluate the basis functions.


\section{Numerical evaluation of classical polylogarithms}
\label{sec:lin}

The classical (or Euler) polylogarithm $\Li_n(x)$ is defined recursively as
\begin{align}
\Li_n(x) &= \int_{0}^x \! \frac{\id y}{y} \, \Li_{n-1}(y)\,,
\end{align}
with
\begin{align}
\Li_1(x) &= -\log ( 1-x ) \,.
\label{eq:li1def}
\end{align}
This implies the relation
\begin{align}
\Li_n(x) &= -G \big( \bar{0}_{n-1}, 1, x \big).
\end{align}
Additionally $\Li_n$ may be expressed as the integral
\begin{align}
\Li_{n}(x)&=\frac{(-1)^{n}}{(n-1)!}\int_0^1 \frac{\log(t)^{n-1}}{(t-1/x)} \id t\,,
\end{align}
which follows from eq. \eqref{eq:GIBP4}.

From eq. \eqref{eq:mpldef} we get the summed expression
\begin{align}
\Li_n(x) &= \sum_{j=1}^{\infty} \frac{x^j}{j^n}\,,
\label{eq:linsum}
\end{align}
which converges whenever $|x| \leq 1$.
When $|x| > 1$, $\Li_n(x)$ may be mapped into the convergent region using the inversion relation \cite{lewin1981}
\begin{align}
\Li_n(x) = (-1)^{n-1} \Li_n \! \left( \tfrac{1}{x} \right) - \tfrac{1}{n!} \log^n(-x) + 2 \sum_{r=1}^{\floor{\tfrac{n}{2}}} \frac{\log^{n-2r}(-x)}{(n-2r)!} \left( 2^{1-2r} - 1 \right) \zeta(2r)
\label{eq:lininversion}
\end{align}
or specifically

\begin{align}
\Li_2(x) &= -\Li_2 \big( \tfrac{1}{x} \big) - \tfrac{1}{2} \log^2(-x) - \tfrac{\pi^2}{6} , \nn \\
\Li_3(x) &= \Li_3 \big( \tfrac{1}{x} \big) - \tfrac{1}{6} \log^3(-x) - \tfrac{\pi^2}{6} \log(-x) , \\
\Li_4(x) &= -\Li_4 \big( \tfrac{1}{x} \big) - \tfrac{1}{24} \log^4(-x) - \tfrac{\pi^2}{12} \log^2(-x) - \tfrac{7 \pi^4}{360} . \!\!\!\!\!\!\!\!\!\! \nn
\end{align}

Close to $|x| = 1$, the sum of eq. \eqref{eq:linsum} converges slowly, making it unsuited for numerical evaluation. An alternative and widely used algorithm was proposed by R. E. Crandall in ref. \cite{Crandall}, and is based on an expansion in the logarithm of the argument of the $\Li_n(x)$, in order to obtain an expression that converges quickly even when $|x| \approx 1$ and $n$ is a small number. The algorithm which we propose in the current section is based on similar considerations. The expansion in question is \cite{Crandall, lewin1981}
\begin{align}
\Li_n \! \left( e^{-\alpha} \right) &= \frac{(-\alpha)^{n-1}}{(n-1)!} \big( H_{n-1} - \log(\alpha) \big) + \!\!\! \sum_{\substack{m=0{}\\m\neq n-1}}^{\infty} \!\! \frac{\zeta (n-m)}{m!} (-\alpha)^m\,,
\label{eq:lincrandall}
\end{align}
where $\alpha=-\log(x)$, $H_n$ denote the harmonic numbers, and $\zeta(x)$ the Riemann zeta function.
We recall that $\zeta(-n) = -B_{n+1}/(n+1)$ for $n \geq 1$, with $B_n$ denoting the Bernoulli numbers.
Specifically this gives
\begin{align}
\Li_2 \! \left( e^{-\alpha} \right) &= \tfrac{\pi^2}{6} - \alpha - \tfrac{1}{4} \alpha^2 + \alpha \log(\alpha) + \sum_{n=1}^{\infty} \tfrac{B_{2n}}{2n(2n+1)!} \alpha^{2n+1}\,, \nn \\
\Li_3 \! \left( e^{-\alpha} \right) &= \zeta_3 - \tfrac{\pi^2}{6} \alpha + \tfrac{3}{4} \alpha^2 + \tfrac{1}{12} \alpha^3 - \tfrac{1}{2} \alpha^2 \log(\alpha) - \sum_{n=1}^{\infty} \tfrac{B_{2n} }{2n(2n+2)!} \alpha^{2n+2}\,, \label{eq:li234crandall} \\
\Li_4 \! \left( e^{-\alpha} \right) &= \tfrac{\pi^4}{90} - \zeta_3 \alpha + \tfrac{\pi^2}{12} \alpha^2 - \tfrac{11}{36} \alpha^3 - \tfrac{1}{48} \alpha^4 + \tfrac{1}{6} \alpha^3 \log(\alpha) + \sum_{n=1}^{\infty} \tfrac{B_{2n} }{2n(2n+3)!} \alpha^{2n+3}\,. \nn
\end{align}

The factorial decay of the terms in eq. \eqref{eq:lincrandall} makes it converge faster than the defining sum of eq. \eqref{eq:linsum}, and for that reason eq. \eqref{eq:lincrandall} is desirable for numerical evaluation. Crandall's algorithm evaluates $\Li_n$ by eq. \eqref{eq:linsum} for $|x|<x_0$  where typically $x_0=1/2$, by eq. \eqref{eq:lininversion} for $|x|>|1/x_0|$, and by eq. \eqref{eq:lincrandall} for $|x_0|\leq x\leq|1/x_0|$. There is, however, an expression with a similar factorial convergence, that expands $\Li_n(1 - e^{-\alpha})$ instead, and which thus is suitable inside the convergent region of eq. \eqref{eq:linsum}. That expression \cite{Devoto:1983tc,Vollinga:2004sn} is 
\begin{align}
\Li_n \! \left( 1-e^{-\alpha} \right) &= \sum_{j=0}^{\infty} \frac{C_{n,j} \alpha^{j+1}}{(j+1)!} 
\label{eq:linbernoulli}
\end{align}
with
\begin{align}
C_{1,j} \; = \; \delta_{j,0} \;\;\;\;\;\;\;\; \tekst{and} \;\;\;\;\;\;\;\; C_{n+1,\,j} \; = \; \sum_{k=0}^{j} \left( \! \begin{array}{c} j \\ k \end{array} \! \right) \frac{B_{j-k}}{k+1} C_{n,k}\,.
\label{eq:linbernoullicoefs}
\end{align}
For $n=2$ this simplifies to
\begin{align}
\Li_2 \! \left( 1-e^{- \alpha} \right) &= \alpha - \tfrac{1}{4}\alpha^2 + \sum_{m=1}^{\infty} \frac{B_{2m} \, \alpha^{2m+1}}{(2m+1)!}\,,
\label{eq:Li2exp}
\end{align}
with no such simplifications for higher $n$. In any case, the coefficients $C_{n,j}$ in \eqref{eq:linbernoullicoefs} can be calculated once and stored in the numerical code, such that recalculating them is not needed when evaluating the $\Li_n$ in a specific point.

\begin{figure}[ht]
\centering
\includegraphics[width=0.55\textwidth]{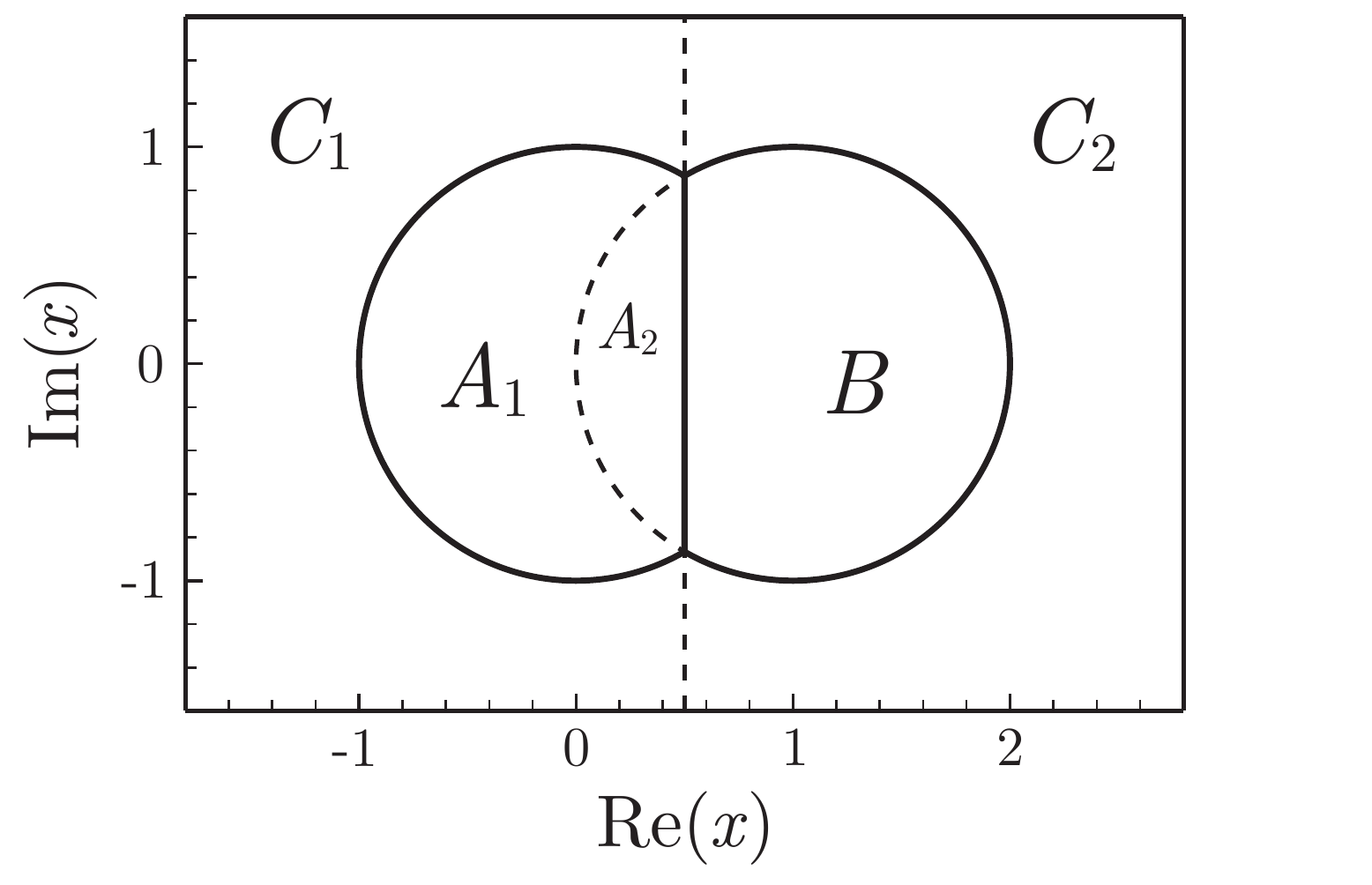}
\caption{This figure shows which expression for $\Li_n(x)$ is used for which values of $x$. In region $A$ we use eq. \eqref{eq:linbernoulli}, in region $B$ we use eq. \eqref{eq:lincrandall}, and in region $C$ we use eq. \eqref{eq:lininversion} to map into region $A$ with $C_1$ mapping to $A_1$ and $C_2$ to $A_2$.}
\label{fig:linregions}
\end{figure}

In our algorithm, we choose to evaluate $\Li_n(x)$ using eqs. \eqref{eq:lincrandall} and \eqref{eq:linbernoulli} together with the inversion formula eq. \eqref{eq:lininversion} to map to the convergent regions. We use eq. \eqref{eq:linbernoulli} whenever $\tekst{Re}(x) \leq \tfrac{1}{2}$ and $|x| \leq 1$, eq. \eqref{eq:lincrandall} whenever $\tekst{Re}(x) > \tfrac{1}{2}$ and $|x-1| \leq 1$, and eq. \eqref{eq:lininversion} combined with eq. \eqref{eq:linbernoulli} otherwise. See fig. \ref{fig:linregions}.

For a detailed discussion of the evaluation of classical polylogarithms and the relations between them, see \cite{lewin1981, Vollinga:2004sn}.


\section{Numerical evaluation of $\Litt(x,y)$}
\label{sec:li22}

$\Litt(x,y)$ is, as can be seen from eq. \eqref{eq:mpldef}, defined as the sum
\begin{align}
\Litt(x,y) &= \! \sum_{i>j>0}^{\infty} \frac{x^i \, y^j}{i^2 \, j^2}\,= \sum_{i=1,\,j=1}^{\infty} \frac{x^i}{(i+j)^2}\frac{(x y)^j}{j^2}\,,
\label{eq:li22fast}
\end{align}
which converges whenever $|x| \leq 1$ and $|xy| \leq 1$.
Additionally it may be expressed as a GPL through eq. \eqref{eq:li22asg}, and as the one-dimensional integral
\begin{align}
\Litt(x,y) &= \int_0^1 \frac{\log(z) \Li_2 (xyz)}{z - \tfrac{1}{x}} \id z\,,
\label{eq:Li22IBP}
\end{align}
which follows from eq. \eqref{eq:GIBP4}.

Outside the region of convergence of eq. \eqref{eq:li22fast} two relations are needed in order to map to the convergent region. One is the stuffle relation (eq. \eqref{eq:stuffle}) which for the case of $\Litt$ becomes
\begin{align}
\Litt(x,y) &= -\Litt(y, x) - \Li_4(xy) + \Li_2(x) \Li_2(y)\,,
\label{eq:li22stuffle}
\end{align}
and which is seen to effectively swap the two arguments. The other needed relation is

\begin{align}
\Litt(x,y) &= \Litt \big( \tfrac{1}{x}, \tfrac{1}{y} \big) - \Li_4(xy) + 3 \Big( \Li_4 \big( \tfrac{1}{x} \big) + \Li_4(y) \Big) + 2 \Big( \Li_3 \big( \tfrac{1}{x} \big) - \Li_3(y) \Big) \log(-xy) \nn \\
& \; + \Li_2 \big( \tfrac{1}{x} \big) \left( \frac{\pi^2}{6} + \frac{\log^2(-xy)}{2} \right) + \frac{1}{2} \Li_2(y) \Big( \log^2(-xy) - \log^2(-x) \Big)\,,
\label{eq:li22inversion}
\end{align}
which is our generalization of the inversion relation eq. \eqref{eq:lininversion} for the case of $\Li_n$, to $\Litt$. As for other similar relations, this inversion relation requires non-zero imaginary parts on $x$, $y$, and $xy$ in order to be guaranteed correct.

\begin{table}
\center
\begin{tabular}{| c | c | l |} \hline
 $|x|$ & $|xy|$ &  \\ \hline \hline
 $<1$ & $<1$ & no mapping needed \\ \hline
 $>1$ & $<1$ & stuffle, eq. \eqref{eq:li22stuffle} \\ \hline
 $>1$ & $>1$ & inversion, eq. \eqref{eq:li22inversion} \\ \hline
 $<1$ & $>1$ & stuffle and inversion \\ \hline
\end{tabular}
\caption{A procedure for mapping $\Litt(x,y)$ to the convergent region. For the case of equalities both cases are in principle applicable.}
\label{tab:li22regions}
\end{table}

That eqs. \eqref{eq:li22stuffle} and \eqref{eq:li22inversion} together can map the whole phase space to the convergent region can be realized from table \ref{tab:li22regions}. That algorithm lays the basis for our implementation of $\Litt$.

Due to the slow convergence of eq. \eqref{eq:li22fast} for values of $|x|$ or $|xy|$ close to one, it is desirable to use another, faster converging expression for these regions in the spirit of eqs. \eqref{eq:linbernoulli} and \eqref{eq:lincrandall} for $\Li_n$.
In ref. \cite{Vollinga:2004sn} eq. \eqref{eq:linbernoulli} is generalized to a larger class of functions, the harmonic polylogarithms, a class which does not include $\Litt$. Yet following steps similar to those of section 4.3 of ref. \cite{Vollinga:2004sn} we derive (see appendix \ref{app:coefficients} for more details)

\begin{align}
\Litt(x,y) &= \sum_{i=0}^{\infty} \sum_{j=1}^{\infty} \alpha^i \beta^j \Big( C^A_{ij} + C^B_{ij} \big( \log(\beta) \! - \! \log(xy) \big) \Big) \label{eq:li22bernoulli} \\
& \; + \sum_{i=1}^{\infty} \alpha^i \bigg( C^C_i \log \! \big( 1 - \tfrac{\beta}{\alpha} \big) + C^D_i \Big( \Li_2 \big( \tfrac{\beta}{\alpha} \big) + \log \! \big( 1 - \tfrac{\beta}{\alpha} \big) \big( \log(\beta) \! - \! \log(xy) \big) \Big) \bigg)\,, \nn
\end{align}
where
\begin{align}
\alpha \; = \; -\log(1-y) \,, \;\;\;\;\;\;\; \beta \; = \; -\log(1-xy) \,,
\end{align}
and where the $C$s denote constants for which the expressions are given in appendix \ref{app:coefficients}, but which may be calculated once and stored, an approach which we use for our implementation.

Also eq. \eqref{eq:lincrandall} may be generalized to the case of $\Litt(x,y)$ (see appendix \ref{app:coefficients} for more details) with the result

\begin{align}
\Litt(x,y) &= \sum_{i=0}^{\infty} \sum_{j=0}^{\infty} \tilde{\beta}^i \tilde{\alpha}^j \Big( K^A_{ij} + K^B_{ij} \log(\tilde{\beta}) + K^C_{ij} \log(\tilde{\beta}+\xi_0) \Big) \nn \\
&\;\; + \sum_{i=0}^{\infty} \tilde{\alpha}^i \Big( K^D_{i} + \tilde{\beta} K^E_{i} \Big) \Big( \log(\tilde{\alpha} + \tilde{\beta}) - \log(\tilde{\alpha} + \tilde{\beta} + \xi_0) \Big) \label{eq:li22crandall} \\
&\;\; + \kur{F}(\tilde{\alpha}, \tilde{\beta}, \xi_0) + \Litt \big( x e^{- \xi_0}, y \big) + \xi_0 \Li_{1,2} \big( x e^{- \xi_0}, y \big)\,, \nn
\end{align}
where
\begin{align}
\tilde{\alpha} \; = \; \log(y) \,, \;\;\;\;\;\;\; \tilde{\beta} \; = \; -\log(xy) \,,
\end{align}
where $\kur{F}(\tilde{\alpha},\tilde{\beta},\xi_0)$ is a rather simple function given in appendix \ref{app:coefficients}. Expressions for the $\xi_0$-dependent coefficients $K$ are likewise given in appendix \ref{app:coefficients}, but note that for a fixed value of $\xi_0$, the coefficients may be calculated once and stored as it was the case for eq. \eqref{eq:li22bernoulli}.
A larger value of the parameter $\xi_0$ decreases the convergence rate of the series expansions in $\tilde{\alpha},\tilde{\beta}$ in \eqref{eq:li22crandall}, while simultaneously increasing the convergence rate of the two MPLs on the last line of eq. \eqref{eq:li22crandall}. We have chosen the fixed value $\xi_0=1$ for our implementation.

\begin{figure}[ht]
\vspace{-0.8cm}
\centering
\includegraphics[width=0.4\textwidth]{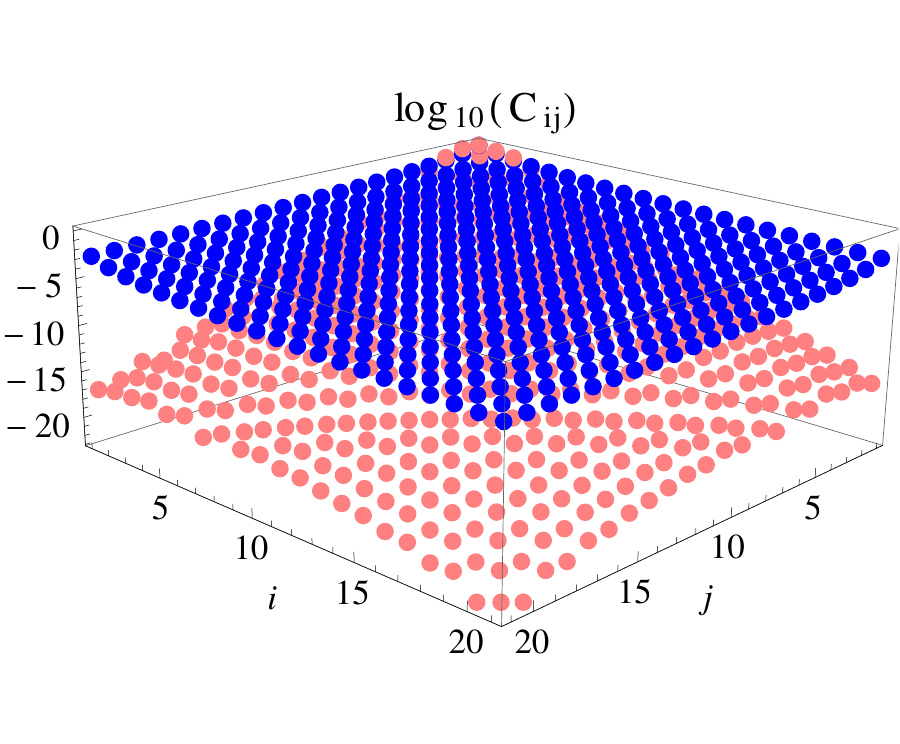}
\vspace{-0.5cm}
\caption{This figure shows the values of the coefficients of eq. \eqref{eq:li22fast} (blue) and eq. \eqref{eq:li22crandall} (red) where the latter is taken to be $K_{ij}^A + K_{ij}^B \log(\tilde{\beta}) + K_{ij}^C \log(\tilde{\beta}+\xi_0)$ evaluated at $\tilde{\beta}=2$ and $\xi_0 = 1$. The faster convergence of eq. \eqref{eq:li22crandall} (as a function of the number of terms) is clearly visible.}
\label{fig:3dplot}
\end{figure}
Unlike the case for $\Li_n$, it is not always preferable to use eqs. \eqref{eq:li22bernoulli} or \eqref{eq:li22crandall} instead of the original defining sum eq. \eqref{eq:li22fast} for reasons of timing. We realize that a precision corresponding to terminating the sum of eq. \eqref{eq:li22fast} at $i=N$ will include around $N^2/2$ terms. But from the factorization property of eq. \eqref{eq:li22fast}, i.e. the fact that it can be written as $\sum_{ij} f(i) g(j)$, one may realize that a proper recursive implementation of the sum will scale\footnote{This is also true for the general $\Li_{m_1,\ldots,m_n}$. A calculation with a cut-off at $i=N$ will include approximately $N^n/(n!)$ terms, but the timing will scale linearly as $nN$.} as only $2N$, i.e. linearly rather than quadratically. This property is not (seemingly) shared by the logarithmic expansions of eqs. \eqref{eq:li22bernoulli} or \eqref{eq:li22crandall}, thus while their faster convergence corresponds to a smaller value of $N$ (as shown in fig. \ref{fig:3dplot}), their timing will still be worse in most cases.

\begin{figure}[ht]
\centering
\includegraphics[width=0.9\textwidth]{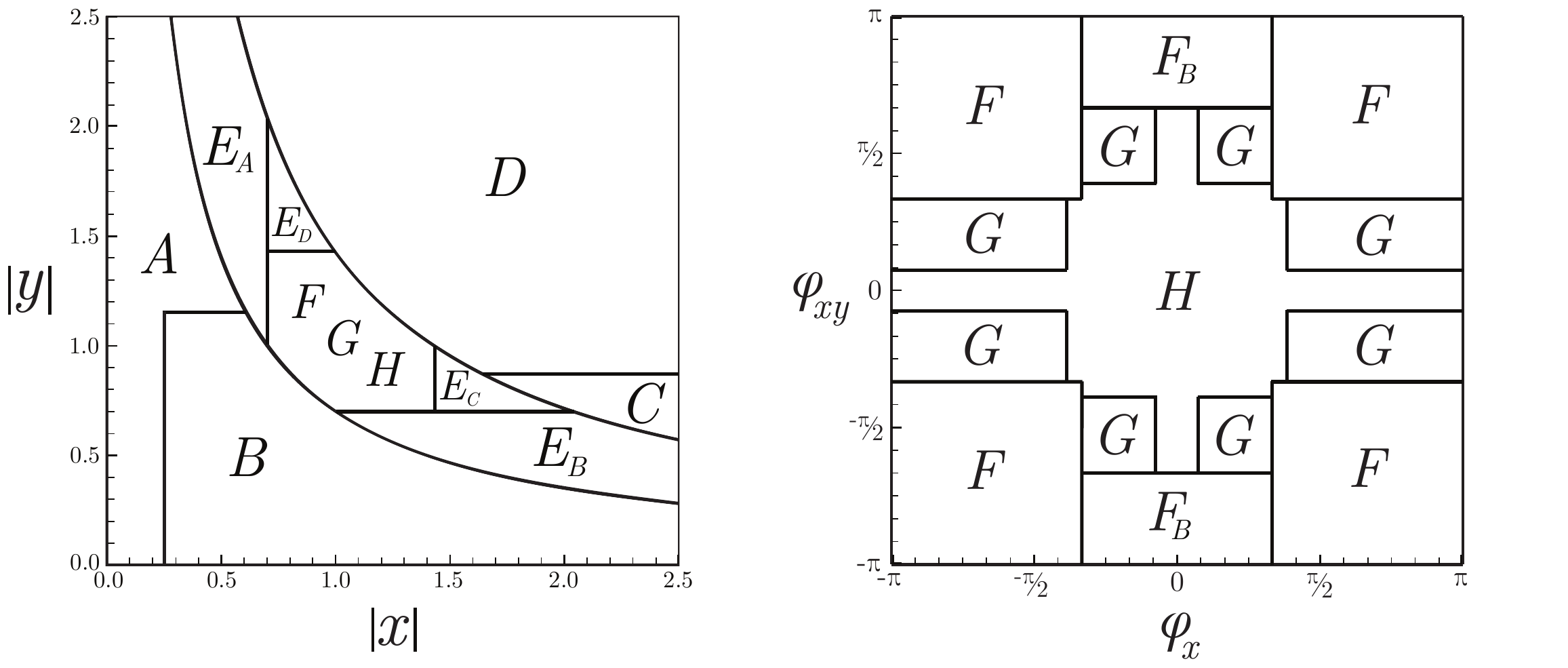}
\caption{This figure is a slightly simplified illustration of the regions into which we split the evaluation of $\Litt(x,y)$ as described in the main text. $\varphi_z$ on the right figure refers to the phase of the complex $z$. Not shown on the figure is the fact that region $C$ is used in place of $D$ whenever $|x|>3.5$.}
\label{fig:li22regions}
\end{figure}

For our implementation we split the evaluation of $\Litt(x,y)$ into a number of regions, as illustrated in fig. \ref{fig:li22regions}. The left figure shows the general case. In region $A$ we use the defining sum eq. \eqref{eq:li22fast}. In region $B$ we combine it with stuffle (eq. \eqref{eq:li22stuffle}), in region $C$ with inversion (eq. \eqref{eq:li22inversion}) and in region $D$ with both. In regions $E$ we apply a resummed version of eq. \eqref{eq:li22fast} in which we analytically perform the sum of the 'diagonal' terms with fixed $i-j$. In the sub-regions it gets combined with $E_A$: nothing, $E_B$: stuffle, $E_C$: inversion, $E_D$: stuffle and inversion.
$FGH$ on the left figure denotes an area in which both $|x|$ and $|xy|$ are close to one. In that case we use the regions shown on the right figure. In regions $F$ we apply the H\"older formula eq. \eqref{eq:holder} with $q=2$ to map to the convergent region (see \cite{Vollinga:2004sn}), and in the sub-region $F_{B}$ this is combined with the stuffle relation. In regions $G$ we apply eq. \eqref{eq:li22bernoulli}, and finally in region $H$, when one of the variables are close to (actual) one, we apply eq. \eqref{eq:li22crandall}. Additionally we choose (everywhere except in regions $A$ and $B$) to use the formulae of app. \ref{app:appA} when applicable.

\subsection*{Considerations on series acceleration}
In order to improve the numerical convergence of $\Litt(x,y)$ for $(|x|\sim1,|y|\sim1)$ we have tried different strategies. For instance we tried to perform the analytic continuation of the series of eq. \eqref{eq:li22fast} from $x=0$ to $x=x_0$, where for example $x_0=-1/2$. We also tried to apply some series acceleration techniques that are explained for example in refs. \cite{1999CoPhC.116.28J,2003math.6302W,2014arXiv1405.2474B}\footnote{One of us (D.T.) would like to thank Dr. E. J. Weniger and Dr. R. Borghi for their kind suggestions and private communication on the topic.}.
Unfortunately, to the best of our knowledge, no transformation that works for our specific problem is known. For any transformation that we attempted the convergence is improved in some specific values of $(x,y)$ but not in others, without any obvious pattern.

For this reason, we decided against using such a procedure, even though it would be interesting to perform more detailed studies in this direction.


\section{The added code}
\label{sec:useofthecode}

Alongside the paper we publish a number of ancillary files containing our implementation of the results from the previous four sections and from appendix \ref{app:integrals}. These are {\tt gtolrules.m}, {\tt gtolexample.nb}, {\tt lievaluate.cpp}, {\tt constants.cpp}, {\tt lievaluateexample.cpp},

\noindent
{\tt lievaluateinterface.tm}, {\tt MakefileLinux}, {\tt MakefileMAC}, and {\tt interfaceexample.nb}.

\subsection*{The reductions}

The file {\tt gtolrules.m} contains the implementation of the relations mapping GPLs to $\log$, $\Li_n$, and $\Litt$ as described in sections \ref{sec:newrelationsGPLs} and \ref{sec:gpl}. 
It has been implemented such that $G(a_1,\ldots,a_n;x)$ have separate relations for any combinations of $a_i=a_j$, $a_i=x$, and $a_i=0$ (the trivial cases of $x=0$ are treated separately). This means that for instance at weight two there are separate relations for
\begin{align}
\begin{array}{ccccc}
G(0,0;x), & G(0,x;x), & G(0,a;x), & G(x,0;x), & G(a,0;x)\,, \\
G(x,x;x), & G(a,a;x), & G(x,a;x), & G(a,x;x), & G(a,b;x)\,.
\end{array}
\end{align}
Whenever $G(a_1,\ldots,a_n;x)$ has $x=a_1$, the result will diverge (unless $x=a_1=1$ and $a_2=\cdots=a_n=0$), as can be seen from the definition of eq. \eqref{eq:gdef} and from the specific expressions in section \ref{sec:gpl}. That divergence is made explicit in the replacement rules by isolating it as powers of the divergent $G(x;x)$ which we express in Mathematica as the symbol {\tt log0}. This is done using the shuffle rules of eq. \eqref{eq:shuffle} as described in ref. \cite{Maitre:2005uu}.
Thus for example the divergent $G(x,a;x)$ is implemented as
\begin{align}
G(x,a;x) \;=\; G(x;x) G(a;x) - G(a,x;x) \;=\; {\tt log0} \log \left( 1 - \tfrac{x}{a} \right) + \Li_2 \big( \tfrac{x}{x-a} \big)\,.
\end{align}
While this treatment of divergences has been applied systematically, there may still be ambiguities with the divergences since no specification of the way the diverging limit is taken is given. We advise the user to take systematic care of any divergences before applying our reductions.

Specifically the reduction is implemented in Mathematica as a set of replacement rules named {\tt gtolrules}. Applying {\tt gtolrules} to an expression containing a symbolic function called {\tt G} of two\footnote{We chose not to implement the trivial $G(;x) = 1$.} to five arguments, will apply the mappings to the {\tt G}s and express the result in terms of the symbolic functions {\tt MyLog}, {\tt MyLi}, and {\tt MyLi22}, and the symbolic functions denoted by {\tt MyT}, {\tt MyP}, {\tt MyR}, {\tt MyH1}, and {\tt MyH2}, which correspond to the various combinations of Heaviside $\theta$-functions described in sections \ref{sec:newrelationsGPLs} and \ref{sec:gpl}. The Heaviside $\theta$-function and $\sgn$-function themselves are denoted by {\tt MyTH} and {\tt MyS} respectively in the replacements rules. Additionally {\tt gtolrules.m} contains the replacement rules {\tt logback} which substitutes {\tt Log} and {\tt PolyLog} for {\tt MyLog} and {\tt MyLi}, and the replacement rules {\tt thetaback} which substitutes the expressions for the $\theta$-functions in.

{\tt gtolrules.m} also contains our results of the $13$ integrals listed in appendix \ref{app:integrals}. They are implemented as the replacement rule {\tt itogrules}, which replace the symbolic function {\tt I1} to {\tt I13} as defined in the appendix, with their expressions in terms of GPLs.

The Mathematica notebook {\tt gtolexample.nb} contains an example of the use of these replacement rules.

\subsection*{The numerical evaluation}

The files {\tt lievaluate.cpp} and {\tt constants.cpp} contain our C++ implementations of $\Li_m(x)$ and $\Litt(x,y)$, described in sections \ref{sec:lin} and \ref{sec:li22}. {\tt lievaluate.cpp} contains mainly functions, and {\tt constants.cpp} a large number of constants needed by the functions. The functions meant for the user are declared as
\begin{verbatim}
complex<double> li1(complex<double> &x);
complex<double> li2(complex<double> &x);
complex<double> li3(complex<double> &x);
complex<double> li4(complex<double> &x);
complex<double> li5(complex<double> &x);
complex<double> li6(complex<double> &x);
complex<double> libasic(int n, complex<double> &x);
complex<double> li22basic(complex<double> &x, complex<double> &y);
complex<double> li(int n, complex<double> &x);
complex<double> li22(complex<double> &x, complex<double> &y);
\end{verbatim}
Many additional functions and constants which are meant for internal use, are also declared in {\tt lievaluate.cpp} and {\tt constants.cpp}, but in order to prevent naming conflicts, they have all been assigned names starting with the letters 'WTF' - the initials of the authors.

Of the functions mentioned above, the first six contain our implementation of $\Li_1(x)$ to $\Li_6(x)$. $\Li_1(x)$ is implemented through the definition eq. \eqref{eq:li1def}, while the others are implemented as described in section \ref{sec:lin}. {\tt libasic} calls these six functions for the corresponding values of $n$, and returns zero and prints a warning for other values. Likewise the function {\tt li22basic} contains our implementation of $\Litt(x,y)$ as described in section \ref{sec:li22}.

For some values of the arguments where the integral of eq. \eqref{eq:gdef} hits a branch-cut, GPLs are not well defined, but need an $\eps$-prescription. We follow the prescription of ref. \cite{Vollinga:2004sn} as explained below eq. \eqref{eq:GIBP4}, which is that $G(a_1,\ldots,a_n;x) \rightarrow G(a_1,\ldots,a_n;x (1 - i \eps))$ in these cases. This corresponds to
\begin{align}
\Li_n(x) \rightarrow \Li_n \big( x (1-i \eps) \big) \;\;\;\;\;\;\;\;\; \tekst{and} \;\;\;\;\;\;\;\;\; \Litt(x,y) \rightarrow \Litt\big( x(1-i \eps), y \big)\,.
\end{align}
In the functions {\tt li} and {\tt li22} we have implemented this prescription, together with a similar regularization which is needed internally for real values of $y$ in $\Litt(x,y)$.

The added file {\tt lievaluateexample.cpp} contains three examples of the use of the code, illustrating how to include it, and how to use the functions. Please note that all the functions call their arguments 'by reference', which implies that calling a function with a number directly (such as {\tt li4(-2.3)}) is invalid. 

The files {\tt lievaluateinterface.tm}, {\tt MakefileLinux} and {\tt MakefileMAC} are used for the Mathematica interface to our code, which is made using the MathLink protocol under the Linux or OS X operating systems. One should use the proper makefile according to the operating system and rename it to {\tt Makefile}. Please follow the instructions which are written as comments in {\tt Makefile} on how to set up and compile the interface. For completion, all the functions mentioned above are included as Mathematica versions.
The file {\tt interfaceexample.nb} contains examples on how to use this interface.

\subsection*{Performance of 'lievaluate'.}

The precision of the {\tt double} format used by C++ allows for $15$-$17$ significant digits. Due to numerical instabilities we do not always attain that level of precision, but of $10000$ evaluations of {\tt li22(x,y)} of random\footnote{Where the absolute values were chosen exponentially distributed between $e^{-10}$ and $e^{10}$.} arguments, none had a relative deviation (i.e. 
$\scalebox{.85}{2}\big|\tfrac{a-b}{a+b}\big|$) from the true value greater than $10^{-13}$. We have also performed many tests in extreme cases, such as cases with arguments very close to zero or one, and found similar levels of agreement.


\section{Discussion}
\label{sec:discussion}

In this paper we presented relations expressing any generalized polylogarithm up to weight four in terms of a basis of logarithms, classical polylogarithms, and the function $\Litt$. Additionally we presented algorithms allowing the evaluation of these functions anywhere in complex phase space.

To express a Feynman integral expressible in terms of GPLs, using such a basis, may be done in general using the symbol map \cite{Goncharov:2010jf, Duhr:2012po} and its co-product based generalization \cite{Duhr:2012fh} in which a minimal basis of functions gets fitted to the 'symbol' of the Feynman integral. The expressions presented in this paper will not in general produce as compact expressions, since the functional basis used in section \ref{sec:gpl} is calculated at the level of the individual GPL, rather than at the level of the entire expression. Additionally our expressions for the reductions have not been fully simplified or optimized with respect to size or computation time.

Yet from the point of view of numerical evaluation of GPLs of general complex variables, we believe our method to be highly competitive. When faced with this problem one previously had three options. A generally inefficient numerical integration, such as the one described below eq. \eqref{eq:GIBP}, a potentially highly demanding reduction using symbols, and an iterated series of algebraic manipulations of the GPLs as implemented for example in {\tt GiNaC} \cite{Vollinga:2004sn}, which in the end potentially requires the evaluation of a vastly increased number of GPLs. 

In conclusion, our expressions should firstly be regarded as a proof of concept that such a reduction is possible in all cases, something which was only conjectured in \cite{Duhr:2012po}, and additionally as an alternative way to obtain numerical expressions for GPLs using only numerical implementations of $\Li_n$ and $\Litt$, such as those described in sections \ref{sec:lin} and \ref{sec:li22} and added as described in section \ref{sec:useofthecode}. 

We regard our over-all approach to the evaluation of $\Litt$, as described in table \ref{tab:li22regions}, to be solid. But the expressions, eqs. \eqref{eq:li22bernoulli} and \eqref{eq:li22crandall} that we use for the evaluation in slowly convergent regions have room for improvement. Perhaps it is possible to resum some of the terms, or reexpress them in a way that allows for a quicker evaluation as it is the case for the defining sum of eq. \eqref{eq:li22fast}. Another option could be to continue the investigation of the series acceleration techniques mentioned at the end of section \ref{sec:li22}.

There is nothing which in principle prevents the approach used in this paper from being extended to higher weights, such as the six which are needed for three-loop calculations. Extending to weight six is conjectured in ref. \cite{Duhr:2012po} to require the extra special functions $\Li_{2,3}$, $\Li_{2,4}$, $\Li_{3,3}$, and $\Li_{2,2,2}$, and they may likely be evaluated using an approach similar to that which we propose for $\Litt$. The reduction to such a set, will require the evaluation of a large number of integrals, similar to higher weight versions of those listed in section \ref{sec:gpl} and appendix \ref{app:integrals}, and it is that step which we believe to be the biggest obstacle for continuing.


\acknowledgments{%
The authors would like to thank Costas Papadopoulos, Thomas Gehrmann, Andreas von Manteuffel, Erik Panzer, Narek Martirosyan, Yang Zhang, Francesco Coradeschi, Ernst Joachim Weniger, and Riccardo Borghi for useful discussions.

In particular we would like to thank Thomas Gehrmann and Costas Papadopoulos for their helpful coments on the draft in its intermediate stages. 

This work was primarily performed at NCSR Demokritos.
CW and DT were primarily supported by the Research Funding Program ARISTEIA, HOCTools (co-financed by the European Union (European Social Fund ESF) and Greek national funds through the Operational Program ``Education and Lifelong Learning'' of the National Strategic Reference Framework (NSRF)).
The work of HF is supported by the European Commission through the HiggsTools Initial Training Network PITN-GA-2012-316704.
}

\appendix


\section{Relations for $\Litt$}
\label{app:appA}

In this appendix we list various relations for $\Litt(x,y)$.

At certain special values, $\Litt$ may be expressed in terms of simpler functions:
\begin{align}
\Litt(0,x) &= 0\,, & \Litt(1,1) &= \frac{\pi^4}{120} = \tfrac{3}{4} \zeta_4\,, \nn \\
\Litt(x,0) &= 0\,, & \Litt(-1,-1) &= -\frac{\pi^4}{480} = -\tfrac{3}{16} \zeta_4\,, \label{eq:A1}
\end{align}
\begin{align}
\Litt(x,x) &= \frac{1}{2} \Big( \Li_2^{\,2}(x) - \Li_4(x^2) \Big)\,, \\
\Litt \big( \tfrac{1}{x},x \big) \! &= 3 \Li_4(x) - \frac{1}{2} \Li_2(x) \Big( \Li_2(x) + \log^2(-x) + \pi^2 \Big) - \frac{\pi^4}{90}\,, \\
\Litt(x,1) &= -2 \Big( \Li_4 \big( \tfrac{1}{1-x} \big) + \Li_4 \big( \tfrac{x}{x-1} \big) + \Li_4 (x) \Big) + 2 \Li_3(x) \log(1-x) \nn \\
& \;\;\; + \frac{1}{2} \Li_2^{\,2}(x) - \frac{1}{6} \log^4(1-x) + \frac{1}{3} \log^3(1-x) \log(-x) \\
& \;\;\; + \frac{\pi^2}{6} \log^2(1-x) - 2 \zeta_3 \log(1-x) + \frac{\pi^4}{45}\,. \nn
\label{eq:Li22_x1}
\end{align}
Note that $\Litt(x,1)$ is equivalent to the harmonic polylogarithm $G(0,1,0,1;x)$. An expression for $\Litt(1,x)$ may be obtained from the above using the stuffle relation of eq. \eqref{eq:li22stuffle}. We note that some of the above relations require nonzero imaginary part of $x$ in order to be well defined and correct.

It is occasionally needed to take limits of $\Litt(x,y)$. When one of the arguments goes to zero and the other remains finite, the function vanishes as per eq. \eqref{eq:A1}. When one of the two arguments diverge and the other stays non-zero, the function diverges and the divergence may be extracted as powers of logarithms using the inversion relations eqs. \eqref{eq:lininversion} and \eqref{eq:li22inversion}. Finally when one argument diverges and the other vanishes, we have the limits
\begin{align}
\lim_{z \rightarrow 0} \Litt(x/z,yz) \, = \, - \Li_4 (xy) \quad\quad \tekst{and} \quad\quad \lim_{z \rightarrow 0} \Litt(xz,y/z) \, = \, 0.
\end{align}

There are also a number of relations for $\Litt$ that are of a general nature. Examples are the stuffle relation eq. \eqref{eq:li22stuffle} and the inversion relation eq. \eqref{eq:li22inversion} mentioned in the main text. Another relation worth mentioning is the duplication relation \cite{Ablinger:2013cf}
\begin{align}
\Litt(x,y) + \Litt(x,-y) + \Litt(-x,y) + \Litt(-x,-y) &= \frac{1}{4} \Litt \big( x^2, y^2 \big)\,.
\end{align}


\section{Details of $\Litt$ expressions}
\label{app:coefficients}

In this appendix we sketch the derivations of eqs. \eqref{eq:li22bernoulli} and \eqref{eq:li22crandall} and list the constants appearing in their expressions.

In order to derive eq \eqref{eq:li22bernoulli}, we start from eq. \eqref{eq:Li22IBP} and apply variable substitutions

\begin{align}
\Litt(x,y) &= \int_0^p \frac{\log \left( \frac{1 - e^{-q}}{xy} \right) \Li_2 \! \left( 1 - e^{-q} \right)}{e^{q-t} - 1} \id q\,,
\end{align}
where
\begin{align}
q = -\log \big( 1-xyz \big) \;,\;\;\; p = -\log(1-xy) \;,\;\;\; t = -\log(1-y)\,. \nn
\end{align}
We expand all terms by the following
\begin{align}
\log \left( \tfrac{1 - e^{-q}}{xy} \right) &= \log\left(\tfrac{1}{xy}\right) + \log(q) + \sum_{i=0}^{\infty} \frac{B_{i+1} q^{i+1}}{(i+1) (i+1)!}\,, \nn \\
\frac{1}{e^{q-t}-1} &= \frac{1}{q-t} + \sum_{i=0}^{\infty} \frac{B_{i+1} (q-t)^i}{(i+1)!}\,, \nn \\
\Li_2 \! \left( 1 - e^{-q} \right) &= \sum_{i=0}^{\infty} \frac{q^{i+1} B_i}{(i+1)!}\,. \label{eq:Bern0}
\end{align}
Note the last expression is equivalent to eq. \eqref{eq:Li2exp}. 
After performing all integrations we obtain eq. \eqref{eq:li22bernoulli}.

The constants of eq. \eqref{eq:li22bernoulli} are

\begin{align}
C^B_{ij} &= \big( 1 - \delta_{j,1} \big) (-1)^i \sum_{\nu=0}^{j-2} \bin{i+\nu}{\nu} \frac{B_{i+\nu+1} \, B_{j-\nu-2}}{j \, (i+\nu+1)! \, (j-\nu-1)!} \;\; + \;\; \frac{B_{i+j-1}}{j \, (i+j)!}\,, \nn \\
C^C_i &= \big( 1 - \delta_{i,1} \big) \sum_{\nu=0}^{i-2} \frac{B_{\nu+1} \, B_{i-\nu-2}}{(\nu+1) \, (\nu+1)! \, (i-\nu-1)!}\,,
\end{align}
\begin{align}
C^A_{ij} &= \big( 1 - \delta_{j,1} - \delta_{j,2} \big) (-1)^i \sum_{\nu=0}^{j-3} \bin{i+\nu}{\nu} \frac{B_{i+\nu+1}}{(i+\nu+1)!} \! \!\sum_{\mu=0}^{j-3-\nu}\!\!\! \frac{B_{\mu+1} \, B_{j-3-\nu-\mu}}{(\mu+1) \, (\mu+1)! \, j \, (j-2-\nu-\mu)!} \nn \\
& \;\; + \big( 1 - \delta_{j,1} \big) (-1)^{i+1} \sum_{\nu=0}^{j-2} \bin{i+\nu}{\nu} \frac{B_{j-\nu-2} \, B_{i+\nu+1}}{j^2 \, (j-\nu-1)! \, (i+\nu+1)!} \nn \\
& \;\; + \big( 1 - \delta_{i,0} \delta_{j,1} \big) \sum_{\nu=0}^{i+j-2} \frac{B_{\nu+1} \, B_{i+j-\nu-2}}{j \, (\nu+1) \, (\nu+1)! \, (i+j-\nu-1)!} \;\; - \;\; \frac{B_{i+j-1}}{j^2 \, (i+j)!}\,,\nn \\
C^D_i &= \frac{B_{i-1}}{i!}\,.\nn
\end{align}

To obtain eq. \eqref{eq:li22crandall}, one splits up the integral \eqref{eq:Li22IBP} in two integrals, namely over the two intervals $[0,\exp(-\xi_0)]$ and $[\exp(-\xi_0),1]$, where $\xi_0$ is an arbitrarily chosen positive number.

\begin{align}
\Litt(x,y) \; = \; \tekst{f}_1 + \tekst{f}_2
\end{align}
where
\begin{align}
\tekst{f}_1 &= \int_0^{\exp(-\xi_0)} \frac{\log(t) \Li_2 \big( txy \big)}{t-1/x} \id t\,, \nn \\
\tekst{f}_2 &= \int_{\exp(-\xi_0)}^1 \frac{\log(t) \Li_2 \big( txy \big)}{t-1/x} \id t \; = \; \int_{\tilde \beta + \xi_0}^{\tilde \beta} \!\!\!\!\! \frac{(\tilde \beta-z) \, \Li_2 \! \big(e^{-z} \big)}{e^{z+\tilde \alpha}-1} \id z\,,
\end{align}
and
\begin{align}
\tilde \beta = -\log(xy) \;,\;\; \tilde \alpha = \log(y) \;,\;\; z = -\log \! \big( txy \big)\,.
\end{align}

In the first integral $\tekst{f}_1$ one rescales the variable as $t\rightarrow t/\exp(-\xi_0)$ and expresses the integral in terms of a sum of an $\Li_{2,2}$ and $\Li_{1,2}$ with argument $(x e^{-\xi_0}, y)$. For $\xi_0$ large enough, one may then safely use the convergent series expressions \eqref{eq:mpldef} in this interval.

For the second integral $\tekst{f}_2$  one applies the following expansion formulas
\begin{equation}
\frac{1}{e^{z+\alpha}-1}  = \frac{1}{z+\alpha} + \sum_{m=0}^{\infty} c_m (z+\alpha)^{m}, \ \ \Li_2 \big( e^{-z} \big)  = z \log(z) + \sum_{n=0}^{\infty} \kappa_n z^n\,, \label{eq:Bern}
\end{equation}
where
\begin{align}
c_m \, = \, \frac{B_{m+1}}{(m+1)!} \qquad\;\; \tekst{and} \qquad\;\; \kappa_0= \frac{\pi^2}{6}\,, \;\;\, \kappa_1=-1\,, \;\;\, \kappa_n=\tfrac{B_{n-1}}{(n-1) n!}\,\,(n\geq2)\,.
\label{eq:candk}
\end{align}
The first equation in eqs. \eqref{eq:Bern} is equivalent to the second equation in eqs. \eqref{eq:Bern0}, while the second equation is the same as the first one in eqs. \eqref{eq:li234crandall}. By computing all resulting integrals one finds eq. \eqref{eq:li22crandall}.

The additive function $\kur{F}(\tilde{\alpha}, \tilde{\beta}, \xi_0)$ is given as

\begin{align}
\kur{F}(\tilde{\alpha}, \tilde{\beta}, \xi_0) &= (\tilde{\alpha}^2 + \tilde{\alpha} \tilde{\beta}) \bigg( \Li_2 \left( -\tfrac{\tilde{\beta} + \xi_0}{\tilde{\alpha}} \right) - \Li_2 \left( \tfrac{-\tilde{\beta}}{\tilde{\alpha}} \right) + \log( \tilde{\beta} + \xi_0 ) \log \left( \tfrac{\tilde{\alpha} + \tilde{\beta} + \xi_0}{\tilde{\alpha}} \right) \\
&\;\;\;\; - \log(\tilde{\beta}) \log \left( \tfrac{\tilde{\alpha} + \tilde{\beta}}{\tilde{\alpha}} \right) \bigg) + 2 \pi i \, \sgn \big( \tekst{Im}(\tilde{\alpha}) \big) (\tilde{\alpha} + \tilde{\beta}) \tilde{\alpha} \log(-\tilde{\alpha}) \, \delta(\tilde{\alpha},\tilde{\beta},\xi_0)\,, \nn
\end{align}
where $\delta(\tilde{\alpha},\tilde{\beta},\xi_0)$ is given as the function
\begin{align}
\delta(\tilde{\alpha},\tilde{\beta},\xi_0) &= \begin{cases} 1 & \tekst{if} \;\; |\arg(\tilde{\beta}+\xi_0)|<|\arg(-\tilde{\alpha})|<|\arg(\tilde{\beta})| \;\;\; \tekst{and} \\
 & \sgn(\arg(\tilde{\beta})) = \sgn(\arg(-\tilde{\alpha})) \;\;\; \tekst{and} \;\;\; 1 < \left| \frac{\tekst{Im}(\tilde{\beta})}{\tekst{Im}(\tilde{\alpha})} \right| \\ 0 & \tekst{otherwise}\,. \end{cases}
\end{align}
For the constants $K$ (which still have dependence on the free parameter $\xi_0$) we use \eqref{eq:candk}
and then

\begin{align}
K^A_{ij} &= \big( 1 - \delta_{i,0} - \delta_{i,1} \big) \frac{1}{i(i-1)} \sum_{\nu=0}^{i-2} \kappa_{\nu} c_{i+j-\nu-2} \bin{i+j-\nu-2}{i-\nu-2}  \\
& \;+ \big( 1 - \delta_{i,0} \big) \sum_{m=j}^{\infty} c_m \bin{m}{m-j} \!\! \sum_{\substack{\nu=\Max(0, \\ m+2-i-j)}}^{\infty} \!\! \bin{i+\nu-1}{i-1} \frac{- \, \xi_0^{\,\nu}}{i-1+\nu} \, \kappa_{\nu+i+j-m-2} \nn \\
& \;+ \sum_{m=j}^{\infty} c_m \bin{m}{m-j} \!\! \sum_{\substack{\nu=\Max(0, \\ m+2-i-j)}}^{\infty} \!\! \bin{i+\nu}{i} \frac{\xi_0^{\,\nu}}{i+\nu} \, \kappa_{\nu+i+j-m-2}\nn \\
& \;+ \big( 1 - \delta_{i,0} \big) \, (-1)^{j+1} \sum_{\nu=1}^{\infty} \kappa_{i+j+\nu-1} \, \frac{\xi_0^{\,\nu}}{\nu+i-1} \bin{i+\nu-1}{i-1} \nn \\
& \;+ \big( 1 - \delta_{j,0} \big) \, (-1)^j \sum_{\nu=1}^{\infty} \kappa_{i+j+\nu-1} \, \frac{\xi_0^{\,\nu}}{\nu+i} \bin{i+\nu}{i} \nn \\
& \;+ \big( 1 - \delta_{i,0} \big) \!\!\!\!\!\!\!\!\! \sum_{\nu=\Max(1, 3-i)}^{\infty} \!\!\!\!\!\! c_{\nu+i+j-3} \bin{\nu+i+j-3}{j} \frac{\xi_0^{\, \nu}}{(\nu+i-1)^2} \bin{\nu+i-1}{i-1} \nn \\
& \;+ \!\!\!\!\! \sum_{\nu=\Max(1, 3-i)}^{\infty} \!\!\!\!\!\! c_{i+j+\nu-3} \bin{i+j+\nu-3}{j} \frac{-\xi_0^{\, \nu}}{(i+\nu)^2} \bin{i+\nu}{i} \nn \\
& \; + \delta_{j,0} \sum_{\nu=1}^{\infty} \kappa_{i+\nu-1} \frac{\xi_0^{\,\nu}}{i+\nu} \bin{i+\nu}{i} \,+\, \delta_{i,0} \delta_{j,0} \frac{- \, \xi_0^{\,2}}{4} \,+\, \delta_{i,1} \delta_{j,0} \frac{\xi_0}{2} \,+\, \delta_{i,0} \delta_{j,1} \xi_0\,, \nn \\
K^B_{ij} &= \big( 1 - \delta_{i,0} - \delta_{i,1} - \delta_{i,2} \big) \, c_{i+j-3} \frac{1}{i(i-1)} \bin{i+j-3}{j} + \delta_{i,2} \delta_{j,0} \frac{1}{2} + \delta_{i,1} \delta_{j,1}\,, \nn \\
K^C_{ij} &=  \big( 1 - \delta_{i,0} \big) \!\!\!\!\!\!\!\!\! \sum_{\nu=\Max(0, 3-i)}^{\infty} \!\!\!\!\!\! c_{\nu+i+j-3} \bin{\nu+i+j-3}{j} \frac{-\xi_0^{\, \nu}}{\nu+i-1} \bin{i-1+\nu}{i-1} \nn \\
& \;\;\; + \!\!\!\!\!\! \sum_{\nu=\Max(0, 3-i)}^{\infty} \!\!\!\!\!\! c_{\nu+i+j-3} \bin{\nu+i+j-3}{j} \frac{\xi_0^{\, \nu}}{\nu+i} \bin{\nu+i}{i}\nn \\
& \;\;\; + \delta_{i,0} \delta_{j,0} \frac{\xi_0^{\,2}}{2} - \delta_{i,0} \delta_{j,1} \xi_0 - \delta_{i,2} \delta_{j,0} \frac{1}{2} - \delta_{i,1} \delta_{j,1}\,, \nn 
\end{align}
\begin{align}
K^D_{i} &= \big( 1 - \delta_{i,0} \big) (-1)^{i+1} \kappa_{i-1}\,, \nn \\
K^E_{i} &= (-1)^i \kappa_i\,. \nn 
\end{align}


\section{Table of integrals}
\label{app:integrals}

The importance of GPLs is due to the fact they allow the integration (at least formally) of a large class of functions, by applying algebraic rules starting from the definition of the GPL, given by eq. (\ref{eq:gdef}). For example

\begin{align}
\int_0^z \frac{G(a_1,\ldots,a_n;t)G(b_1,\ldots,b_n;t)}{(t-c)(t-d)}\,dt
\end{align}

can easily be integrated in terms of GPLs by applying the shuffle rule of eq. \eqref{eq:shuffle} on the numerator and partial fractioning on the denominator. Furthermore for both of the following integrals
\begin{align}
\int_0^z \frac{G(a_1,\ldots,a_n;t)}{(t-b)^2}\,dt \qquad \mbox{or}\qquad \int_0^z t^3 G(a_1,\ldots,a_n;t)\,dt
\end{align}
the integration can be performed by integration by parts. In practice any combination of GPLs and rational functions can be formally integrated by the use of GPLs.

Here we provide an example, where we can formally integrate a large class of integrals containing logarithms (assuming that the $b_i$ are all different)
\begin{align}
\int_0^z & \frac{\left(\log(1-t/a_1)+\log(1-t/a_2)+\ldots\log(1-t/a_m)\right)^n}{(t-b_1)(t-b_2)\ldots(t-b_k)}\,dt=\nn\\
=&\int_0^z \left(\sum_{q_1+q_2+\ldots+q_m=n}\!\!\!\!\!\!\!\!n!\frac{G(a_1;t)^{q_1}}{q_1!}\frac{G(a_2;t)^{q_2}}{q_2!}\ldots \frac{G(a_m;t)^{q_m}}{q_m!}\right)\!\!\!\left(\sum_{i=1}^k\frac{1}{(t-b_i)}\frac{1}{\prod_{j\neq i}(b_i-b_j)}\right)dt\nn\\
=& \, n! \sum_{i=1}^k\left(\sum_{\overline T}\frac{G(b_i,\overline T_n(\{a_1,\ldots,a_m\});z)}{\prod_{j\neq i}(b_i-b_j)}\right)\,,
\end{align}
where the \emph{tuple} $\overline T_n(\{a_1,\ldots,a_m\})$ is the set of all possible vectors of dimension $n$, made up of all possible combinations and orders of the elements ${a_1,\ldots,a_n}$. For example $$\overline T_2(\{a,b,c\})=\{\{a, a\}, \{a, b\}, \{a, c\}, \{b, a\}, \{b, b\}, \{b, c\}, \{c, a\}, \{c, b\}, \{c, c\}\}\,.$$

If the result of the integral is a combinations of GPLs of weights $\leq 4$, it is now possible to explicitly express it in terms of the basis functions by using the results of this paper.
Here we list some of the most useful integrals, of which some have been known for a long time \cite{lewin1981,Devoto:1983tc}. We do however explicitly perform the analytic continuation and extend the integrals to the full complex plane of the variables. Furthermore we extend the list of integrals to more complicated cases, including cases with results involving $\Litt$, which to the best of our knowledge are generally unknown. In order to obtain expressions in terms of $\log$, $\Li_n$, and $\Litt$, all that is needed it to combine the expressions of this appendix with the reductions described in sections \ref{sec:newrelationsGPLs} and \ref{sec:gpl}.
The expressions in this appendix are included in an auxiliary file, as explained in section \ref{sec:useofthecode}.
\begin{align}
I_1(z;a,b)&=\int_0^z \frac{\log(t) \log(1-t/a)}{t-b} \,dt \,=\, G(b,0,a;z)+G(b,a,0;z) \\
&= \; G(0,z)G(b,a;z)-G(0,b,a;z)\,,\nn\\
I_2(z;a,b)&=\int_0^z \frac{\log(1-t/a) \log(1-t/b)}{t} \,dt \,=\, G(0,a,b;z)+G(0,b,a;z)\,,\\
I_3(z;a,b,c)&=\int_0^z \frac{\log(1-t/a) \log(1-t/b)}{t-c} \,dt \,=\, G(c,a,b;z)+G(c,b,a;z)\,.
\end{align}
From now on, in order to keep the list of integrals relatively short, we will include only the most general cases without considering null letters or letters equaling each other, since they can be easily obtained with a similar strategy.
\begin{align}
I_4(z;a,b) &= \int_0^z \log(1-t/a) \log(1-t/b) \,dt \\
&= 2z + (z-a) (G(a, b;z) - G(a;z)) + (z-b) (G(b, a;z) - G(b;z))\,,\nn\\
I_5(z;a,b,c) &= \int_0^z \log(1-t/a) \log(1-t/b)\log(1-t/c) \, dt \\
&= -6z + (z-a) \Big( 2 G(a;z)\! -\! G(a, b;z)\! -\! G(a, c;z)\! +\! G(a, b, c;z)\! +\! G(a,c,b;z) \Big)\nn\\
&\;\;\; +(z-b) \Big( 2G(b;z)-G(b, a;z) - G(b, c;z) + G(b, a, c;z) + G(b, c, a;z) \Big)\nn\\
&\;\,\; +(z-c) \Big( 2 G(c;z) - G(c, a;z) -G(c, b;z) + G(c, a, b;z) + G(c, b, a;z) \Big)\,,\nn\\
I_{6}(z;a,b,c,d) &= \int_0^z \log(1-t/a) \log(1-t/b)\log(1-t/c)\log(1-t/d) \,dt\\
&= (z-a) \Big( -6 G(a;z) + 2 \big( G(a, b;z) + G(a, c;z) + G(a, d;z) \big) - G(a, b, c;z)\nn\\
&\;\;\;\;\; - G(a, b, d;z) - G(a, c, b;z) - G(a, c, d;z) - G(a, d, b;z) - G(a, d, c;z)\nn\\
&\;\;\;\;\; + G(a, b, c, d;z) + G(a, b, d, c;z) + G(a, c, b, d;z) + G(a, c, d, b;z)\nn\\
&\;\;\;\;\; +  G(a, d, b, c;z) + G(a, d, c, b;z) \Big) + (\text{cyclic in}~\{a,b,c,d\}) + 24 z\,, \nn \\
I_7(z;a,b,c) &= \int_0^z \frac{\log(1-t/a)^2 \log(1-t/b)}{t-c} \,dt \\
&= 2 \Big( G(c, a, a, b;z) + G(c, a, b, a;z) + G(c, b, a, a;z) \Big)\,,\nn\\
I_8(z;a,b,c,d) &= \int_0^z \frac{\log(1-t/a) \log(1-t/b)\log(1-t/c)}{t-d} \,dt \\
&= G(d, a, b, c;z) + G(d, a, c, b;z) + G(d, b, a, c;z) +  G(d, b, c, a;z)\nn\\
& \;\;\;\; + G(d, c, a, b;z) + G(d, c, b, a;z)\,,\nn \\
I_{9}(z;a,b,c)&=\int_0^z \frac{\Li_2(t/a) \log(1-t/b)}{t-c} \,dt \\
&=-G(c, 0, a, b;z) - G(c, 0, b, a;z) - G(c, b, 0, a;z)\,,\nn
\end{align}
\begin{align}
I_{11}(z;a,b) &= \int_0^z \,\Li_2(t/a) \log(1-t/b)^2 \,dt \\
&= \!2(z-a) \Big( 3 G(a;z) - 2 G(a, b;z) + G(a, b, b;z) \Big) + 2z \Big(-6 - G(0, a;z)\nn\\
&\;\;\;\; + G(0, a, b;z) + G(0, b, a;z) - G(0, a, b, b;z)- G(0, b, a, b;z) - G(0, b, b, a;z) \Big)\nn\\
&\;\; +2(z-b) \Big( 3 G(b;z) - 2 G(b, a;z) - G(b, b;z) + G(b, 0, a;z) + G(b, a, b;z)\nn\\
&\;\;\;\; + G(b, b, a;z) - G(b, 0, a, b;z) - G(b, 0, b, a;z) - G(b, b, 0, a;z) \Big)\,,\nn \\
I_{10}(z;a,b,c) &= \int_0^z \,\Li_2(t/a) \log(1-t/b)\log(1-t/c) \,dt \\
&= (z-a) \Big( 6 G(a;z) - 2 G(a, b;z) - 2 G(a, c;z) + G(a, b, c;z) + G(a, c, b;z) \Big)\nn\\
&\;\;\; + z \Big(-12 - 2 G(0, a;z) + G(0, a, b;z) + G(0, a, c;z) +  G(0, b, a;z)\nn\\
&\;\;\;\;\; + G(0, c, a;z) - G(0, a, b, c;z) -  G(0, a, c, b;z) - G(0, b, a, c;z)\nn\\
&\;\;\;\;\; - G(0, b, c, a;z)- G(0, c, a, b;z) - G(0, c, b, a;z) \Big)\nn\\
&\;\;\; + (z-b) \Big( 3 G(b;z) - 2 G(b, a;z) - G(b, c;z) + G(b, 0, a;z) + G(b, a, c;z)\nn\\
&\;\;\;\;\; + G(b, c, a;z) -  G(b, 0, a, c;z) - G(b, 0, c, a;z)- G(b, c, 0, a;z) \Big)\nn\\
&\;\;\; + (z-c) \Big( 3 G(c;z) - 2 G(c, a;z) - G(c, b;z) + G(c, 0, a;z) + G(c, a, b;z)\nn\\
&\;\;\;\;\; + G(c, b, a;z) - G(c, 0, a, b;z) - G(c, 0, b, a;z) - G(c, b, 0, a;z) \Big)\,,\nn \\
I_{12}(z;a,b) &= \int_0^z \,\Li_2(t/a) \Li_2(t/b) \,dt \\
&= (z-a) \Big(-3 G(a;z) + 2 G(a, b;z) - G(a, 0, b;z) \Big)\nn\\
&\;\;\; + (z-b) \Big(-3 G(b;z) + 2 G(b, a;z) - G(b, 0, a;z) \Big)\nn\\
&\;\;\; + z \Big(6 + G(0, a;z) + G(0, b;z) - 2 G(0, a, b;z) -  2 G(0, b, a;z) \nn \\
&\;\;\;\,\; + 2 G(0, 0, a, b;z) + 2 G(0, 0, b, a;z) + G(0, a, 0, b;z) + G(0, b, 0, a;z) \Big)\,,\nn \\
I_{13}(z;a,b) &= \int_0^z \,\Li_3(t/a) \log(1-t/b) \,dt \\
&=\! (z-a) \Big(3 G(a;z) - G(a, b;z) \Big) + z \Big( -4 - 2 G(0, a;z) + G(0, 0, a;z)\nn\\
&\;\;\;\; +\! G(0, a, b;z)\! +\! G(0, b, a;z)\! -\! G(0, 0, a, b;z)\! -\! G(0, 0, b, a;z)\! -\! G(0, b, 0, a;z) \Big)\nn\\
&\;\; +(z-b) \Big( G(b;z) - G(b, a;z) + G(b, 0, a;z) - G(b, 0, 0, a;z) \Big)\,.\nn
\end{align}


\section{Series expansions of hypergeometric functions}
\label{app:hypergeometric}

GPLs can be used to obtain, in some cases, a series expansion of hypergeometric functions in a purely algebraic way.
Alternative methods are already available as explained in \cite{Huber:2005yg,Kalmykov:2007pf,Huang:2012qz,Bytev:2013gva}.
Here we consider an different approach, inspired by the integration technique used in \cite{Papadopoulos:2014lla},
which is in turn inspired by well established integration techniques, see for instance \cite{Catani:1996vz,Binoth:2000ps}.
Let us consider as an example the following Euler's expression
\begin{equation}
B(2+\epsilon,3+2\epsilon)~{}_2 F_1(\epsilon,2+\epsilon;5+3\epsilon;z)=\int_0^1 x^{1+\epsilon}(1-x)^{2+2\epsilon}(1-z x)^{-\epsilon}\,dx\,,
\label{eq:B2F1def}
\end{equation}
where $B(b,c)$ is the beta function, ${}_2 F_1$ is the Gauss hypergeometric and $|z|\leq1$. In practice one may require some expansion in terms of the small parameter $\epsilon$ of the integral in eq. \eqref{eq:B2F1def}.

The procedure is as follows. The first step is to perform the following expansions
\begin{align}
& x^{n+\epsilon}=x^n e^{\epsilon\log(x)}=x^n \sum_{i=0}^\infty \epsilon^i\frac{\log(x)^i}{i!} =x^n \sum_{i=0}^\infty \epsilon^i G(\overline 0_i;x)\,, \label{eq:subst1}\\
& (a-x)^{n+\epsilon}=(a-x)^n a^\epsilon (1-x/a)^{\epsilon}=(a-x)^n a^\epsilon \sum_{i=0}^\infty \epsilon^i G(\overline a_i;x)\,.
\label{eq:subst2}
\end{align}
Note that the first expression eq. (\ref{eq:subst1}) is valid for any complex $x$, while the second eq. (\ref{eq:subst2}) is in general valid only for real $x$ (but still allows for a complex parameter $a$). Generally this is not a problem, since $x$ is the integration variable as in eq. (\ref{eq:B2F1def}).

As a second step we can substitute the expansion for each term in the integrand in eq. (\ref{eq:B2F1def})
\begin{align}
&\int_0^1 x(1-x)^2\left(\sum_{i=0}^\infty\epsilon^i G(\overline 0_i;x)\right)\left(\sum_{j=0}^\infty 2^j\epsilon^j G(\overline 1_j;x)\right)\left(\sum_{k=0}^\infty (-\epsilon)^k G(\overline{(1/z)}_k;x)\right) dx\,.
\label{eq:B2F1sub}
\end{align}

The third step is to truncate the sums at the desired order, work out the products and apply the shuffle algebra. For example we may compute the result up to $\epsilon^2$
\begin{align}
\int_0^1& (x-1)^2 x \Big(1+  \epsilon  (G(0;x) + 2 G(1;x) - G(1/z;x)) + \epsilon^2 (G(0, 0;x) + 2 G(0, 1;x)\nn\\
&- G(0, 1/z;x) + 2 G(1, 0;x) + 4 G(1, 1;x) - 2 G(1, 1/z;x) - G(1/z, 0;x)\nn\\
&- 2 G(1/z, 1;x) + G(1/z, 1/z;x) +\kur{O}(\epsilon^3)\Big) dx\,.
\label{eq:B2F1expand}
\end{align}

At this point the integration is performed term by term via integration by parts and applying the definition eq. (\ref{eq:gdef}).

In principle this technique can be applied to any integrand involving an arbitrarily complicated product of $n$ terms, and the expansion can be computed up to arbitrary order in $\epsilon$.
However there are three main issues.
First of all the method cannot be generally applied to arbitrary powers. For example a term $x^{1/2+\epsilon}$ will generally lead to non-rational functions and then the result cannot be expressed in terms of GPLs. A second problem can arise in the case of nested integrations over different variables. For example, after integrating eq. (\ref{eq:B2F1expand}) the resulting GPLs will be dependent on the variable $z$ in their letters as $G(\ldots,f(z),\ldots;1)$. At a second integration step we may want to integrate over the variable $z$. It is possible to rewrite any GPL in standard form $G(\ldots;z)$ for any rational functions in z (at least in principle), but it is not always possible for non-rational expressions (see \cite{Panzer:2014caa}). Actually both these problems are related to the fact that not all hypergeometric functions can be expanded as combinations of GPLs, but require more general functions.

The third problem is related to possible divergences. This problem can be solved by subtracting the main divergent part as follows (see \cite{Papadopoulos:2014lla}). Let's consider for simplicity an integrand with a divergence in $x=0$ as for example $x^{-1+\epsilon}f(x)$ where $f(x)$ is regular at $x=0$. Then
\begin{align}
\int_0^1 x^{-1+\epsilon}f(x)=\int_0^1 x^{-1+\epsilon}f(0)+\int_0^1 x^{-1+\epsilon}(f(x)-f(0))\,.
\end{align}
Since $f(0)$ is just a constant, the first integral on the right can be integrated analytically, while the second is regularized since $f(x)-f(0)$ is not only finite, but goes to zero at least linearly in $x$. For example we can consider the following integration
\begin{align}
\int_0^x& \frac{t^{-1+\epsilon}(a-t)^\epsilon}{\epsilon(1-\epsilon)} ~dt=\frac{1}{\epsilon(1-\epsilon)}\left[a^\epsilon\int_0^x t^{-1+\epsilon} ~dt + \int_0^x t^{-1+\epsilon}((a-t)^\epsilon-a^\epsilon) ~dt\right]\label{eq:subtraction1}\\
&=\frac{1}{\epsilon(1-\epsilon)}\left[\frac{a^\epsilon x^\epsilon}{\epsilon}+\int_0^x \left(\frac{\sum_{i=0}^\infty\epsilon^i G(\overline 0_i;t)}{t}\right)\Bigg(a^\epsilon\sum_{j=1}^\infty\epsilon^j G(\overline a_j;t)\Bigg) ~dt\right]\label{eq:subtraction2}\\
&=\frac{a^\epsilon}{\epsilon(1-\epsilon)}\left[\frac{x^\epsilon}{\epsilon}+ \!\! \sum_{i=0,~j=1}^\infty \!\! \epsilon^{i+j}\Bigg(\sum_{\overline c_k\in\overline 0_i \sha \overline a_j} \! G(0,\overline c_k;x)\Bigg)\right].\label{eq:subtraction3}
\end{align}
Note that the right sum in eq. (\ref{eq:subtraction2}) has lower limit one, due to the procedure of subtracting the divergence.

In case the previously explained problems do not appear, the procedure presented in this appendix is suitable for an automatized implementation in a computer code. Furthermore, for expansions including GPLs up to weight 4, we can obtain analytic results as explained in sects.  \ref{sec:newrelationsGPLs} and \ref{sec:gpl}.


\section{Specific expressions for GPLs}
\label{app:w3}

This appendix contains the explicit expressions\footnote{In case of 3 letters, these can always be mapped to 0,1,-1 and the r.h.s. will contain HPLs of rescaled letters (partly together with $x$). We thank our referee for pointing this out.} for all GPLs at weight 3. In principle, expressions for these GPLs can be partially obtained for example by the code Mathematica, but once different scales are present the results will not be in a form which is valid everywhere, as the full analytic structure will be absent. The results presented in this appendix are obtained using the methods described in sections \ref{sec:newrelationsGPLs} and \ref{sec:gpl}. The function $\text{S}(a,b,c,0;1)$ that appears in eq. \eqref{eq:Gabc0prel} and which only depends on GPLs up to weight 3 is given at the end of this appendix.
For the definitions of the functions $\text{T},\mathcal{H}_1$ and $\text{P}$ appearing in the expressions see sects. \ref{sec:newrelationsGPLs} and \ref{sec:gpl}.

\begin{align}
\scl{G(0,0,0;x)} &\scl{\;\;=\;\;} \scl{\tfrac{1}{6} \log(x)^3\,,} \\
\scl{G(0,0,a;x)} &\scl{\;\;=\;\;} \scl{-\Li_3\left(\tfrac{x}{a}\right)\,,} \\
\scl{G(0,a,0;x)} &\scl{\;\;=\;\;} \scl{2 \Li_3\left(\tfrac{x}{a}\right)-\log(x) \Li_2\left(\tfrac{x}{a}\right)\,,} \\
\scl{G(a,0,0;x)} &\scl{\;\;=\;\;} \scl{\log(x) \Li_2\left(\tfrac{x}{a}\right)-\Li_3\left(\tfrac{x}{a}\right)+\tfrac{1}{2} \log(x)^2 \log\left(1-\tfrac{x}{a}\right)\,,} \\
\scl{G(0,a,a;x)} &\scl{\;\;=\;\;} \scl{\log\left(1-\tfrac{x}{a}\right) \Li_2\left(1-\tfrac{x}{a}\right)-\Li_3\left(1-\tfrac{x}{a}\right)+\tfrac{1}{2} \log\left(\tfrac{x}{a}\right) \log\left(1-\tfrac{x}{a}\right)^2+\zeta_3\,,} \\
\scl{G(a,0,a;x)} &\scl{\;\;=\;\;} \scl{-\log\left(1-\tfrac{x}{a}\right) \Li_2\left(\tfrac{x}{a}\right)-2 \log\left(1-\tfrac{x}{a}\right) \Li_2\left(1-\tfrac{x}{a}\right)+2 \Li_3\left(1-\tfrac{x}{a}\right) -\log\left(\tfrac{x}{a}\right) \log\left(1-\tfrac{x}{a}\right)^2-2 \zeta_3\,,} \\
\scl{G(a,a,0;x)} &\scl{\;\;=\;\;} \scl{\log\left(1-\tfrac{x}{a}\right) \Li_2\left(\tfrac{x}{a}\right)+\log\left(1-\tfrac{x}{a}\right) \Li_2\left(1-\tfrac{x}{a}\right)-\Li_3\left(1-\tfrac{x}{a}\right) +\tfrac{1}{2} \log(x) \log\left(1-\tfrac{x}{a}\right)^2+\tfrac{1}{2} \log\left(\tfrac{x}{a}\right) \log\left(1-\tfrac{x}{a}\right)^2+\zeta_3\,,} \\
\scl{G(a,a,a;x)} &\scl{\;\;=\;\;} \scl{\tfrac{1}{6} \log\left(1-\tfrac{x}{a}\right)^3\,,}
\end{align}
\begin{align}
\scl{G(0,a,b;x)} &\scl{\;\;=\;\;} \scl{i \pi \log \left(x \left(\tfrac{1}{a}-\tfrac{1}{b}\right)\right)^2 \text{sgn}\left(\text{Im} \! \left(\tfrac{b}{x}\right)\right) \mathcal{H}_1\left(\tfrac{a}{x},\tfrac{b}{x}\right)+i \pi \log\left(1-\tfrac{a}{b}\right)^2 \text{sgn}\left(\text{Im}\left(\tfrac{a}{b}\right)\right) \text{T}\left(1,1 \! - \! \tfrac{x}{b},1 \! - \! \tfrac{a}{b}\right)-\log\left(1-\tfrac{x}{b}\right) \Li_2\left(\tfrac{x-b}{a-b}\right) } \\
& \;\;\; \scl{+\log\left(1-\tfrac{x}{b}\right) \Li_2\left(\tfrac{a (x-b)}{x (a-b)}\right) +\Li_3\left(\tfrac{x-b}{a-b}\right)-\Li_3\left(\tfrac{a (x-b)}{x (a-b)}\right)-\Li_3\left(\tfrac{b}{b-a}\right)+\Li_3\left(\tfrac{x}{a}\right)-\log\left(1-\tfrac{x}{b}\right) \Li_2\left(1-\tfrac{b}{x}\right)+\Li_3\left(1-\tfrac{b}{x}\right)-\tfrac{1}{6} \log\left(\tfrac{a b}{a x-b x}\right)^3 } \nn \\
& \;\;\; \scl{-\tfrac{1}{2} \log\left(1-\tfrac{x}{b}\right)^2 \log\left(\tfrac{a-x}{a-b}\right) +\tfrac{1}{2} \log\left(1-\tfrac{x}{b}\right)^2 \log\left(\tfrac{b (a-x)}{x (a-b)}\right)-\tfrac{1}{6} \pi ^2 \log\left(\tfrac{a b}{a x-b x}\right)+\tfrac{1}{6} \log\left(\tfrac{b}{x}\right)^3-\tfrac{1}{2} \log\left(1-\tfrac{x}{b}\right)^2 \log\left(\tfrac{b}{x}\right)+\tfrac{1}{6} \pi ^2 \log\left(\tfrac{b}{x}\right)\,,} \nn
\end{align}
\begin{align}
\scl{G(a,0,b;x)} &\scl{\;\;=\;\;} \scl{-i \pi \log\left(\tfrac{a-x}{b}\right)^2 \text{sgn}\left(\text{Im}\left(\tfrac{b}{x}\right)\right) \mathcal{H}_1\left(\tfrac{b}{a},\tfrac{b}{x}\right)-i \pi \log\left(1-\tfrac{x}{a}\right)^2 \text{sgn}\left(\text{Im}\left(\tfrac{x}{a}\right)\right) \text{T}\left(1,1-\tfrac{x}{b},1-\tfrac{x}{a}\right)-\log\left(1-\tfrac{x}{a}\right) \Li_2\left(\tfrac{a}{b}\right)} \\
&\;\;\; \scl{-\log\left(1-\tfrac{x}{b}\right) \Li_2\left(\tfrac{b-x}{a-x}\right)+\log\left(1-\tfrac{x}{b}\right) \Li_2\left(\tfrac{a (b-x)}{b (a-x)}\right)+\Li_3\left(\tfrac{b-x}{a-x}\right)-\Li_3\left(\tfrac{a (b-x)}{b (a-x)}\right)-\Li_3\left(\tfrac{a}{b}\right)+\Li_3\left(\tfrac{a}{a-x}\right)+\log\left(1-\tfrac{x}{b}\right) \Li_2\left(1-\tfrac{b}{x}\right)} \nn \\
&\;\;\; \scl{-\Li_3\left(1-\tfrac{b}{x}\right)-\Li_3\left(\tfrac{x}{b}\right)+\tfrac{1}{6} \log\left(-\tfrac{b}{a-x}\right)^3+\tfrac{1}{6} \pi ^2 \log\left(-\tfrac{b}{a-x}\right)-\tfrac{1}{2} \log\left(1-\tfrac{x}{b}\right)^2 \log\left(\tfrac{a-b}{a-x}\right)+\tfrac{1}{2} \log\left(1-\tfrac{x}{b}\right)^2 \log\left(\tfrac{x (a-b)}{b (a-x)}\right)-\tfrac{1}{6} \log\left(\tfrac{b}{x}\right)^3} \nn\\
&\;\;\; \scl{+\tfrac{1}{2} \log\left(\tfrac{b}{x}\right) \log\left(1-\tfrac{x}{b}\right)^2-\tfrac{1}{6} \pi ^2 \log\left(\tfrac{b}{x}\right)\,,} \nn
\end{align}
\begin{align}
\scl{G(a,b,0;x)} &\scl{\;\;=\;\;} \scl{i \pi \Big( \text{sgn}\left(\text{Im}\left(\tfrac{b}{x}\right)\right)[\log\left(\tfrac{a-x}{b}\right)^2  \mathcal{H}_1\left(\tfrac{b}{a},\tfrac{b}{x}\right)-\log\left(x \left(\tfrac{1}{a}-\tfrac{1}{b}\right)\right)^2 \mathcal{H}_1\left(\tfrac{a}{x},\tfrac{b}{x}\right)]-\log\left(1-\tfrac{a}{b}\right)^2 \text{sgn}\left(\text{Im}\left(\tfrac{a}{b}\right)\right) \text{T}\left(1,1-\tfrac{x}{b},1-\tfrac{a}{b}\right)} \\
&\;\;\; \scl{+2 \log(x) \log\left(1-\tfrac{a}{b}\right) \text{sgn}\left(\text{Im}\left(\tfrac{b}{x}\right)\right) \text{T}\left(1,1-\tfrac{x}{b},1-\tfrac{a}{b}\right)+\log\left(1-\tfrac{x}{a}\right)^2 \text{sgn}\left(\text{Im}\left(\tfrac{x}{a}\right)\right) \text{T}\left(1,1-\tfrac{x}{b},1-\tfrac{x}{a}\right)\Big) -\log(x) \left(\Li_2\left(\tfrac{b}{b-a}\right)-\Li_2\left(\tfrac{x-b}{a-b}\right)\right)} \nn \\
&\;\;\; \scl{+\log\left(1-\tfrac{x}{a}\right) \Li_2\left(\tfrac{a}{b}\right)+\log\left(1-\tfrac{x}{b}\right) \Li_2\left(\tfrac{b-x}{a-x}\right)-\log\left(1-\tfrac{x}{b}\right) \Li_2\left(\tfrac{a (b-x)}{b (a-x)}\right)+\log\left(1-\tfrac{x}{b}\right) \Li_2\left(\tfrac{x-b}{a-b}\right)-\log\left(1-\tfrac{x}{b}\right) \Li_2\left(\tfrac{a (x-b)}{x (a-b)}\right)-\Li_3\left(\tfrac{b-x}{a-x}\right)} \nn \\
&\;\;\; \scl{+\Li_3\left(\tfrac{a (b-x)}{b (a-x)}\right)-\Li_3\left(\tfrac{x-b}{a-b}\right)+\Li_3\left(\tfrac{a (x-b)}{x (a-b)}\right)+\Li_3\left(\tfrac{a}{b}\right)+\Li_3\left(\tfrac{b}{b-a}\right)-\Li_3\left(\tfrac{a}{a-x}\right)-\Li_3\left(\tfrac{x}{a}\right)+\Li_3\left(\tfrac{x}{b}\right)-\tfrac{1}{6} \log\left(-\tfrac{b}{a-x}\right)^3+\tfrac{1}{6} \log\left(\tfrac{a b}{a x-b x}\right)^3} \nn \\
&\;\;\; \scl{+\tfrac{1}{2} \log\left(1-\tfrac{x}{b}\right)^2 \log\left(\tfrac{a-b}{a-x}\right)+\tfrac{1}{2} \log\left(1-\tfrac{x}{b}\right)^2 \log\left(\tfrac{a-x}{a-b}\right)-\tfrac{1}{2} \log\left(1-\tfrac{x}{b}\right)^2 \log\left(\tfrac{b (a-x)}{x (a-b)}\right)-\tfrac{1}{2} \log\left(1-\tfrac{x}{b}\right)^2 \log\left(\tfrac{x (a-b)}{b (a-x)}\right)} \nn \\
&\;\;\; \scl{+\log(x) \log\left(1-\tfrac{x}{b}\right) \log\left(\tfrac{a-x}{a-b}\right)-\pi ^2 \left(\tfrac{1}{6} \log\left(-\tfrac{b}{a-x}\right)-\tfrac{1}{6} \log\left(\tfrac{a b}{a x-b x}\right)\right)\,,}\nn
\end{align}
\begin{align}
\scl{G(a,b,b;x)} &\scl{\;\;=\;\;} \scl{-i \pi \log\left(1-\tfrac{a}{b}\right)^2 \text{sgn}\left(\text{Im}\left(\tfrac{a}{b}\right)\right) \text{T}\left(1,1-\tfrac{x}{b},1-\tfrac{a}{b}\right)+\log\left(1-\tfrac{x}{b}\right) \Li_2\left(\tfrac{x-b}{a-b}\right)-\Li_3\left(\tfrac{x-b}{a-b}\right)+\Li_3\left(-\tfrac{b}{a-b}\right)} \\
&\;\;\; \scl{+\tfrac{1}{2} \log\left(1-\tfrac{x}{b}\right)^2 \log\left(\tfrac{a-x}{a-b}\right)\,,}\nn
\end{align}
\begin{align}
\scl{G(a,b,a;x)} &\scl{\;\;=\;\;} \scl{2 i \pi  \log\left(1-\tfrac{b}{a}\right)^2 \text{sgn}\left(\text{Im}\left(\tfrac{b}{a}\right)\right) \text{T}\left(1,1-\tfrac{x}{a},1-\tfrac{b}{a}\right)+2 i \pi  \log\left(1-\tfrac{b}{a}\right) \log\left(1-\tfrac{x}{a}\right) \text{sgn}\left(\text{Im}\left(\tfrac{a}{x}\right)\right) \text{T}\left(1,1-\tfrac{x}{a},1-\tfrac{b}{a}\right)} \\
&\;\;\; \scl{-\log\left(1-\tfrac{x}{a}\right) \Li_2\left(\tfrac{a}{a-b}\right)-\log\left(1-\tfrac{x}{a}\right) \Li_2\left(\tfrac{a-x}{a-b}\right)+2 \Li_3\left(\tfrac{a-x}{a-b}\right)-2 \Li_3\left(\tfrac{a}{a-b}\right)\,,} \nn
\end{align}
\begin{align}
\scl{G(a,a,b;x)} &\scl{\;\;=\;\;} \scl{-i \pi \log\left(1-\tfrac{b}{a}\right)^2 \text{sgn}\left(\text{Im}\left(\tfrac{b}{a}\right)\right) \text{T}\left(1,1-\tfrac{x}{a},1-\tfrac{b}{a}\right)-2 i \pi  \log\left(1-\tfrac{b}{a}\right) \log\left(1-\tfrac{x}{a}\right) \text{sgn}\left(\text{Im}\left(\tfrac{a}{x}\right)\right) \text{T}\left(1,1-\tfrac{x}{a},1-\tfrac{b}{a}\right)} \\
&\;\;\; \scl{+\log\left(1-\tfrac{x}{a}\right) \Li_2\left(\tfrac{a}{a-b}\right)-\Li_3\left(\tfrac{a-x}{a-b}\right)+\Li_3\left(\tfrac{a}{a-b}\right)-\tfrac{1}{2} \log\left(1-\tfrac{x}{a}\right)^2 \log\left(\tfrac{x-b}{a-b}\right)+\tfrac{1}{2} \log\left(1-\tfrac{x}{a}\right)^2 \log\left(1-\tfrac{x}{b}\right)\,,} \nn
\end{align}
\begin{align}
\scl{G(a,b,c;x)} &\scl{\;\;=\;\;} \scl{\tfrac{1}{6} \log\left(\tfrac{c-b}{a}\right)^3-\tfrac{1}{6} \log\left(\tfrac{(a-c) (c-b)}{(a-b) c}\right)^3-\tfrac{1}{6} \log\left(\tfrac{c-b}{a-x}\right)^3+\tfrac{1}{6} \log\left(\tfrac{(a-c) (c-b)}{(a-b) (c-x)}\right)^3-\tfrac{1}{2} \log\left(\tfrac{a}{a-b}\right) \log\left(\tfrac{b}{b-c}\right)^2+\Li_2\left(\tfrac{x-b}{a-b}\right) \log\left(\tfrac{b-x}{b-c}\right)} \\
&\;\;\; \scl{-\tfrac{1}{2} \log\left(1-\tfrac{b}{a}\right) \log\left(\tfrac{b}{b-c}\right)^2 +\tfrac{1}{2} \log\left(\tfrac{a-b}{a-x}\right) \log\left(\tfrac{b-x}{b-c}\right)^2+\tfrac{1}{2} \log\left(\tfrac{a-x}{a-b}\right) \log\left(\tfrac{b-x}{b-c}\right)^2+\Li_3\left(\tfrac{b}{a}\right)+\Li_3\left(\tfrac{b}{b-a}\right)-\Li_3\left(\tfrac{b (a-c)}{a (b-c)}\right)-\Li_3\left(\tfrac{b (a-c)}{(a-b) c}\right)} \nn \\
&\;\;\; \scl{-\Li_3\left(-\tfrac{c}{b-c}\right)+\Li_3\left(1-\tfrac{c}{a}\right) -\Li_3\left(\tfrac{a-c}{a-x}\right)-\Li_3\left(\tfrac{b-x}{a-x}\right)+\Li_3\left(\tfrac{(a-c) (b-x)}{(b-c) (a-x)}\right)+\Li_3\left(\tfrac{(a-c) (b-x)}{(a-b) (c-x)}\right)-\Li_3\left(\tfrac{x-b}{a-b}\right)-\Li_3\left(\tfrac{x-c}{a-c}\right)+\Li_3\left(\tfrac{x-c}{b-c}\right)} \nn \\
&\;\;\; \scl{-\Li_2\left(\tfrac{a-c}{b-c}\right) \log\left(\tfrac{a}{a-c}\right)+\Li_2\left(\tfrac{-c}{b-c}\right) \log\left(\tfrac{a}{a-c}\right)-\Li_2\left(\tfrac{b}{a}\right) \log\left(\tfrac{b}{b-c}\right)-\Li_2\left(\tfrac{b}{b-a}\right) \log\left(\tfrac{b}{b-c}\right)+\Li_2\left(\tfrac{b (a-c)}{a (b-c)}\right) \log\left(\tfrac{b}{b-c}\right)+\Li_2\left(\tfrac{b (a-c)}{(a-b) c}\right) \log\left(\tfrac{b}{b-c}\right)} \nn \\
&\;\;\; \scl{+\tfrac{1}{2} \log\left(\tfrac{b}{b-c}\right)^2 \log\left(\tfrac{(b-a) c}{a (b-c)}\right)+\tfrac{1}{2} \log\left(\tfrac{b}{b-c}\right)^2 \log\left(\tfrac{a (c-b)}{(a-b) c}\right)+\Li_2\left(\tfrac{a-c}{b-c}\right) \log\left(\tfrac{a-x}{a-c}\right)-\Li_2\left(-\tfrac{c}{b-c}\right) \log\left(\tfrac{a-x}{a-c}\right)+\Li_2\left(\tfrac{b-x}{a-x}\right) \log\left(\tfrac{b-x}{b-c}\right)} \nn \\
&\;\;\; \scl{-\Li_2\left(\tfrac{(a-c) (b-x)}{(b-c) (a-x)}\right) \log\left(\tfrac{b-x}{b-c}\right)-\Li_2\left(\tfrac{(a-c) (b-x)}{(a-b) (c-x)}\right) \log\left(\tfrac{b-x}{b-c}\right)-\tfrac{1}{2} \log\left(\tfrac{b-x}{b-c}\right)^2 \log\left(-\tfrac{(b-c) (a-x)}{(a-b) (c-x)}\right)-\tfrac{1}{2} \log\left(\tfrac{b-x}{b-c}\right)^2 \log\left(\tfrac{(b-a) (c-x)}{(b-c) (a-x)}\right)+\Li_3\left(-\tfrac{c}{a-c}\right)} \nn \\
&\;\;\; \scl{-\Li_2\left(\tfrac{c-b}{a-b}\right) \log\left(1-\tfrac{x}{c}\right)+\Li_2\left(\tfrac{x-b}{a-b}\right) \log\left(1-\tfrac{x}{c}\right)+\log\left(\tfrac{a-x}{a-b}\right) \log\left(\tfrac{b-x}{b-c}\right) \log\left(1-\tfrac{x}{c}\right)-\pi ^2 \Big(-\tfrac{1}{6} \log\left(\tfrac{c-b}{a}\right)+\tfrac{1}{6} \log\left(\tfrac{(a-c) (c-b)}{(a-b) c}\right)+\tfrac{1}{6} \log\left(\tfrac{c-b}{a-x}\right)} \nn \\
&\;\;\; \scl{-\tfrac{1}{6} \log\left(\tfrac{(a-c) (c-b)}{(a-b) (c-x)}\right)+4 \log\left(1-\tfrac{b}{c}\right) \text{T}\left(1,1-\tfrac{x}{c},1-\tfrac{b}{c}\right) \text{T}\left(\text{P}\left(\tfrac{b}{c},1-\tfrac{x}{c}\right),1-\tfrac{x}{c},1-\tfrac{a}{c}\right) \Big)+i \pi \Big(\text{T}\left(1,\tfrac{b}{b-c},\tfrac{a}{a-c}\right) \text{sgn}\left(\text{Im}\left(\tfrac{c}{a-c}\right)\right) \log\left(\tfrac{a}{a-c}\right)^2} \nn \\
&\;\;\; \scl{+\log\left(\tfrac{b-a}{b-c}\right)^2 \text{T}\left(1,\tfrac{b}{b-c},\tfrac{b-a}{b-c}\right) \text{sgn}\left(\text{Im}\left(\tfrac{a-c}{b-c}\right)\right)-\log\left(\tfrac{b-a}{b-c}\right)^2 \text{T}\left(1,\tfrac{b-x}{b-c},\tfrac{b-a}{b-c}\right) \text{sgn}\left(\text{Im}\left(\tfrac{a-c}{b-c}\right)\right)-2 \Li_2\left(\tfrac{a-c}{b-c}\right) \text{T}\left(1,1-\tfrac{x}{c},1-\tfrac{a}{c}\right) \text{sgn}\left(\text{Im}\left(\tfrac{a}{c}\right)\right)} \nn \\
&\;\;\; \scl{+2 \Li_2\left(-\tfrac{c}{b-c}\right) \text{T}\left(1,1-\tfrac{x}{c},1-\tfrac{a}{c}\right) \text{sgn}\left(\text{Im}\left(\tfrac{a}{c}\right)\right)-2 \log\left(1-\tfrac{a}{c}\right) \log\left(\tfrac{b-a}{b-c}\right) \text{T}\left(1,1-\tfrac{x}{c},1-\tfrac{a}{c}\right) \text{sgn}\left(\text{Im}\left(\tfrac{a}{c}\right)\right)} \nn \\
&\;\;\; \scl{-\mathcal{H}_1\left(1-\tfrac{a}{c},1-\tfrac{b}{c}\right) \log\left(\tfrac{(b-a) c}{(a-c) (c-b)}\right)^2 \text{sgn}\left(\text{Im}\left(\tfrac{b}{c}\right)\right)+2 \log\left(1-\tfrac{b}{c}\right) \log\left(\tfrac{a-b}{a-c}\right) \text{T}\left(1,1-\tfrac{x}{c},1-\tfrac{b}{c}\right) \text{sgn}\left(\text{Im}\left(\tfrac{b}{c}\right)\right)} \nn \\
&\;\;\; \scl{-2 \log\left(1-\tfrac{b}{c}\right) \log\left(\tfrac{a-x}{a-c}\right) \text{T}\left(1,1-\tfrac{x}{c},1-\tfrac{b}{c}\right) \text{sgn}\left(\text{Im}\left(\tfrac{b}{c}\right)\right) +\mathcal{H}_1\left(\tfrac{b-c}{a-c},\tfrac{c-b}{c-x}\right) \log\left(\tfrac{a-x}{b-c}\right)^2 \text{sgn}\left(\text{Im}\left(\tfrac{c-b}{c-x}\right)\right)} \nn \\
&\;\;\; \scl{-\mathcal{H}_1\left(\tfrac{c-a}{c-x},\tfrac{c-b}{c-x}\right) \log\left(-\tfrac{(a-b) (c-x)}{(a-c) (c-b)}\right)^2 \text{sgn}\left(\text{Im}\left(\tfrac{c-b}{c-x}\right)\right)+2 \log\left(\tfrac{b-a}{b-c}\right) \log\left(1-\tfrac{x}{c}\right) \text{T}\left(1,\tfrac{b-x}{b-c},\tfrac{b-a}{b-c}\right) \text{sgn}\left(\text{Im}\left(\tfrac{c-b}{c-x}\right)\right)} \nn \\
&\;\;\; \scl{+\log\left(\tfrac{a-x}{a-c}\right)^2 \text{T}\left(1,\tfrac{b-x}{b-c},\tfrac{a-x}{a-c}\right) \text{sgn}\left(\text{Im}\left(\tfrac{x-c}{a-c}\right)\right)+\mathcal{H}_1\left(\tfrac{b-c}{a-c},1-\tfrac{b}{c}\right) \log\left(\tfrac{a}{b-c}\right)^2 \text{sgn}\left(\text{Im}\left(\tfrac{b}{c}\right)\right)\Big)\,.} \nn
\end{align}

Furthermore with stuffle and shuffle identities one finds the relations
\begin{eqnarray}
\scl{\Li_{3,1}(a,b)} &\scl{\! =\!} & \scl{\tfrac{1}{2}\! \left(\Li_2(b) \Li_2(a b)\! +\! \Li_{2,2}(b,a)\! -\! \Li_{2,2}\left(a b,\tfrac{1}{a}\right)\! \right)-\Li_4(a b)+\log(1-b) (\Li_3(a b)-\Li_3(a))\,,} \nn\\
\scl{\Li_{2,2}(a,b)} &\scl{\! =\!} & \scl{\Li_4(a b)+\log (1-a) \Li_3(b)-\log (1-a) \Li_3(a b)-\Li_{3,1}(a,b)+\Li_{3,1}(b,a)-\Li_{3,1}\left(a b,\frac{1}{b}\right)\,.} \label{eq:Li31repl}
\end{eqnarray}

The function $\text{S}(a,b,c,0;1)$ that appears in eq. \eqref{eq:Gabc0prel} equals
\begin{align}
\scl{\text{S}(a,b,c,0;1)}&\scl{\;\;=\;\;} \scl{-\tfrac{1}{2}G\left(0,\tfrac{a}{c};1\right) \left(G\left(0,\tfrac{1}{c};1\right)+ G\left(a,\tfrac{a}{b};1\right)\right)+\tfrac{1}{2}G\left(0,\tfrac{b}{c};1\right) \left(- G\left(0,\tfrac{b}{a c};1\right)+ G(a,b;1)+ G\left(a,\tfrac{a}{c};1\right)\right)+\tfrac{1}{2} G(0,b;1) \left(G\left(0,\tfrac{b}{a};1\right)+G(a,c;1)\right)} \nn\\
&\;\;\; \scl{+G\left(0,\tfrac{a}{b};1\right) \left(\tfrac{1}{4} G(0,a;1)-\tfrac{1}{2} G\left(0,\tfrac{1}{b};1\right)+\tfrac{1}{2} G(a,0;1)-\tfrac{1}{2} G\left(a,\tfrac{a}{c};1\right)\right)+G\left(0,\tfrac{c}{b};1\right) \left(-\tfrac{1}{2} G\left(0,\tfrac{c}{a b};1\right)+\tfrac{1}{2} G\left(a,\tfrac{a}{b};1\right)+\tfrac{1}{2} G(a,c;1)\right)} \nn\\
&\;\;\; \scl{+G\left(a,\tfrac{a}{b};1\right) \left(\tfrac{1}{2} G\left(\tfrac{c}{b},0;1\right)-\tfrac{1}{2} G\left(\tfrac{a}{c},0;1\right)\right)+G(a,0;1) \left(-G\left(\tfrac{c}{b},0;1\right)+\tfrac{1}{2} G\left(\tfrac{c}{b},\tfrac{a}{b};1\right)+\tfrac{1}{2} G\left(\tfrac{c}{b},c;1\right)+\tfrac{1}{2} G(b,c;1)-\tfrac{1}{2} G\left(\tfrac{a}{c},\tfrac{a}{b};1\right)\right)} \nn\\
&\;\;\; \scl{+G(a,c;1) \left(\tfrac{1}{2} G\left(\tfrac{c}{b},0;1\right)+\tfrac{1}{2} G(b,0;1)\right)+G(0,a;1) \left(G\left(0,\tfrac{c}{b};1\right)-\tfrac{1}{2} G\left(0,\tfrac{a c}{b};1\right)-\tfrac{3}{4} G(0,b;1) +\tfrac{1}{4} G\left(0,\tfrac{a}{c};1\right)+G\left(0,\tfrac{b}{c};1\right)-\tfrac{1}{2} G\left(0,\tfrac{a b}{c};1\right)\right.}\nn\\
&\;\;\; \scl{\left. -\tfrac{3}{4} G(0,c;1)-\tfrac{1}{2} G\left(\tfrac{a}{b},\tfrac{a}{c};1\right)+\tfrac{1}{2} G\left(\tfrac{c}{b},\tfrac{a}{b};1\right)+\tfrac{1}{2} G\left(\tfrac{c}{b},c;1\right)+\tfrac{1}{2} G(b,c;1)-\tfrac{1}{2} G\left(\tfrac{a}{c},\tfrac{a}{b};1\right)+\tfrac{1}{2} G\left(\tfrac{b}{c},b;1\right)+\tfrac{1}{2} G\left(\tfrac{b}{c},\tfrac{a}{c};1\right) +\tfrac{1}{2} G(c,b;1)\right)}\nn\\
&\;\;\; \scl{+G(0,c;1) \left(\tfrac{1}{2} G\left(0,\tfrac{c}{a};1\right)-\tfrac{1}{2} G(a,0;1)-\tfrac{1}{2} G(a,b;1)-G\left(\tfrac{b}{c},\tfrac{a}{c};1\right)\right)+G(0,0,a;1)\left(G\left(\tfrac{b}{a};1\right)+G\left(\tfrac{c}{a};1\right)+G\left(\tfrac{a c}{b};1\right)-G\left(\tfrac{c}{a b};1\right) \right.}\nn\\
&\;\;\; \scl{\left.-G\left(\tfrac{1}{c};1\right)+G\left(\tfrac{a b}{c};1\right)-G\left(\tfrac{b}{a c};1\right)\right)+G\left(\tfrac{1}{b};1\right) \left(G\left(0,0,\tfrac{a}{b};1\right)-G(0,0,a;1)\right)+G\left(\tfrac{c}{a b};1\right) G\left(0,0,\tfrac{c}{b};1\right)-G\left(\tfrac{a b}{c};1\right) G\left(0,0,\tfrac{c}{b};1\right)} \nn\\
&\;\;\; \scl{-G\left(\tfrac{b}{a};1\right) G(0,0,b;1)+G\left(\tfrac{1}{c};1\right) G\left(0,0,\tfrac{a}{c};1\right)-G\left(\tfrac{a c}{b};1\right) G\left(0,0,\tfrac{b}{c};1\right)+G\left(\tfrac{b}{a c};1\right) G\left(0,0,\tfrac{b}{c};1\right)-G\left(\tfrac{c}{a};1\right) G(0,0,c;1)} \nn\\
&\;\;\; \scl{+G\left(\tfrac{b}{c};1\right) \left(-\tfrac{1}{2} G(c;1) G\left(0,\tfrac{a}{c};1\right)-G(0,0,a;1)-\tfrac{1}{2} G(0,a,b;1)-\tfrac{1}{2} G\left(0,a,\tfrac{a}{c};1\right)\right)+G\left(\tfrac{a}{b};1\right) \left(\tfrac{1}{2} G\left(\tfrac{b}{c};1\right) G(0,b;1)+\tfrac{1}{2} G(0,0,a;1) \right.} \nn\\
&\;\;\; \scl{\left. -\tfrac{1}{2} G(0,0,b;1)+\tfrac{1}{2} G\left(0,a,\tfrac{a}{c};1\right)-\tfrac{1}{2} G\left(0,b,\tfrac{b}{c};1\right)\right)+G(c;1) \left(\tfrac{1}{2} G(0,0,a;1)-\tfrac{1}{2} G\left(0,0,\tfrac{a}{c};1\right)-\tfrac{1}{2} G(0,a,b;1)+\tfrac{1}{2} G\left(0,\tfrac{a}{c},\tfrac{b}{c};1\right)\right)} \nn\\
&\;\;\; \scl{+G\left(\tfrac{c}{b};1\right) \left(\tfrac{1}{2} G\left(\tfrac{a}{c};1\right) G(0,c;1)-G(0,0,a;1)-\tfrac{1}{2} G\left(0,a,\tfrac{a}{b};1\right)-\tfrac{1}{2} G(0,a,c;1)+G(a,0,0;1)-\tfrac{1}{2} G\left(a,0,\tfrac{a}{b};1\right)-\tfrac{1}{2} G(a,0,c;1) \right.} \nn\\
&\;\;\; \scl{\left. -\tfrac{1}{2} G\left(a,\tfrac{a}{b},0;1\right)-\tfrac{1}{2} G(a,c,0;1)\right)+G(b;1) \left(-\tfrac{1}{2} G\left(\tfrac{c}{b};1\right) G\left(0,\tfrac{a}{b};1\right)+\tfrac{1}{2} G(0,0,a;1)-\tfrac{1}{2} G\left(0,0,\tfrac{a}{b};1\right)-\tfrac{1}{2} G(0,a,c;1)\right.} \nn\\
&\;\;\; \scl{\left. +\tfrac{1}{2} G\left(0,\tfrac{a}{b},\tfrac{c}{b};1\right)+\tfrac{1}{2} G\left(0,\tfrac{c}{b},\tfrac{a}{b};1\right)-\tfrac{1}{2} G(a,0,c;1)-\tfrac{1}{2} G(a,c,0;1)+\tfrac{1}{2} G\left(\tfrac{c}{b},0,\tfrac{a}{b};1\right)\right)+G\left(\tfrac{a}{c};1\right) \left(\tfrac{1}{2} G(0,0,a;1)-\tfrac{1}{2} G(0,0,c;1) \right.}\nn\\
&\;\;\; \scl{\left. +\tfrac{1}{2} G\left(0,a,\tfrac{a}{b};1\right)-\tfrac{1}{2} G\left(0,\tfrac{c}{b},c;1\right)-\tfrac{1}{2} G\left(0,c,\tfrac{c}{b};1\right)+\tfrac{1}{2} G\left(a,0,\tfrac{a}{b};1\right)+\tfrac{1}{2} G\left(a,\tfrac{a}{b},0;1\right)-\tfrac{1}{2} G\left(\tfrac{c}{b},0,c;1\right)\right)+G(a;1) \left(-\tfrac{1}{2} G\left(0,0,\tfrac{a}{b};1\right) \right.}\nn\\
&\;\;\; \scl{\left. +\tfrac{3}{2} G(0,0,c;1)-\tfrac{1}{2} G\left(0,\tfrac{a}{b},0;1\right)+\tfrac{1}{2} G\left(0,\tfrac{a}{b},\tfrac{a}{c};1\right)-\tfrac{1}{2} G\left(0,\tfrac{c}{b},\tfrac{a}{b};1\right)-\tfrac{1}{2} G\left(0,\tfrac{c}{b},c;1\right)-\tfrac{1}{2} G(0,b,c;1)+\tfrac{1}{2} G\left(0,\tfrac{a}{c},\tfrac{a}{b};1\right)-\tfrac{1}{2} G\left(0,\tfrac{b}{c},b;1\right) \right.}\nn\\
&\;\;\; \scl{\left. -\tfrac{1}{2} G\left(0,\tfrac{b}{c},\tfrac{a}{c};1\right)+\tfrac{1}{2} G(0,c,0;1)+\tfrac{1}{2} G(0,c,b;1)+G\left(\tfrac{c}{b},0,0;1\right)-\tfrac{1}{2} G\left(\tfrac{c}{b},0,\tfrac{a}{b};1\right)-\tfrac{1}{2} G\left(\tfrac{c}{b},0,c;1\right)-\tfrac{1}{2} G\left(\tfrac{c}{b},\tfrac{a}{b},0;1\right)-\tfrac{1}{2} G\left(\tfrac{c}{b},c,0;1\right) \right.} \nn\\
&\;\;\; \scl{\left. -\tfrac{1}{2} G(b,0,c;1)-\tfrac{1}{2} G(b,c,0;1)+\tfrac{1}{2} G\left(\tfrac{a}{c},0,\tfrac{a}{b};1\right)+\tfrac{1}{2} G\left(\tfrac{a}{c},\tfrac{a}{b},0;1\right)\right)\,.} \label{eq:G3abc0}
\end{align}

For the complete expressions at weight four, see the attached file {\tt gtolrules.m}.

\bibliographystyle{ieeetr}
\bibliography{biblio}

\end{document}